%% file: main.tex
\documentclass[preprint,12pt,nopreprintline]{elsarticle}

\usepackage[T1]{fontenc}
\usepackage{lmodern}
\usepackage{mathtools}
\usepackage{amssymb}
\usepackage{booktabs}
\input{sections/table_layout_rev1}
\usepackage{array}
\usepackage{multirow}
\usepackage{float}
\usepackage{needspace}
\usepackage{placeins}
\usepackage{rotating}
\usepackage{pdflscape}
\usepackage{xcolor}
\usepackage{microtype}
\usepackage[hidelinks]{hyperref}

\newif\ifshowrevisions
\showrevisionsfalse
\DeclareRobustCommand{\rev}[1]{\ifshowrevisions\textcolor{blue}{#1}\else#1\fi}
\newenvironment{revision}{\begingroup\ifshowrevisions\color{blue}\fi}{\endgroup}

\mathtoolsset{showonlyrefs}
\setkeys{Gin}{draft=false}
\graphicspath{{figures/}}

\begin{document}

\noeqref{eq:profile-Fh,eq:vmax-plugin,eq:gamma-plugin,eq:profile-covariance,eq:cluster-gaussian-score}

\begin{frontmatter}

\title{Variance-Aware Estimation and Inference for Michaelis--Menten Regression with Heteroscedastic and Clustered Enzyme-Kinetic Data}

\author[aff1]{Mijeong Kim}
\author[aff1]{Minkyoung Cha}
\author[aff1]{Ah Young Jeong\corref{cor1}}

\cortext[cor1]{Correspondence: \href{mailto:berry\_sot@ewha.ac.kr}{berry\_sot@ewha.ac.kr} (Ah Young Jeong).}
\address[aff1]{Department of Statistics, Ewha Womans University, Seoul 03760, South Korea}

\begin{abstract}
\small
\setlength{\baselineskip}{13pt}
Michaelis--Menten regression remains central to enzyme-kinetic analysis, yet routine applications still often rely on homoscedastic nonlinear least squares even when assay variability increases with substrate concentration or repeated readings are clustered.
\rev{We develop original-scale Michaelis--Menten estimation and inference using semiparametric score equations for independent curves and a working-covariance extension for clustered assays.}
In simulation, heteroscedastic modeling improved variance recovery and \rev{shortened intervals in some scenarios} relative to homoscedastic nonlinear least squares, while cluster-aware \rev{profile} and NLME fits \rev{improved confidence-interval coverage for the maximum reaction velocity relative to} pooled analyses that ignored clustering\rev{, with interval calibration depending on the number of independent clusters}.
\rev{We apply the proposed method to self-driving laboratory and soil exoenzyme data.}
The companion \rev{R package} \texttt{inferMM}\rev{, available on CRAN,} implements the resulting workflow for single-curve, grouped, and clustered Michaelis--Menten analysis.
\rev{\texttt{inferMM} supports user-defined, fixed positive variance functions for single-curve and grouped estimation and inference. For independent single curves, it also accommodates non-Gaussian errors.}
These results show that simple variance-function and covariance modeling can stabilize enzyme-kinetic inference when variability changes with substrate concentration or measurements are clustered.
\end{abstract}

\begin{keyword}
\small
clustered measurements \sep heteroscedastic nonlinear regression \sep Michaelis--Menten kinetics \sep semiparametric inference
\end{keyword}

\end{frontmatter}

\input{sections/introduction_rev1}
\input{sections/methods_rev1}

\input{sections/simulation_design_rev1}

Performance measures were computed separately for each data-generating scenario and each competing estimator. For each scenario, we evaluated point estimation, interval estimation, variance-function recovery, and computational performance. The measures used in the comparison are described below.

Let $\theta$ denote either $V_{\max}$ or $K_m$, and let $\hat{\theta}_r$ be the estimate obtained from the $r$th successful replication, $r=1,\ldots,R$, where $R$ is the number of successful fits under a given scenario and method. Point estimation performance was assessed by bias and root mean squared error (RMSE),
\begin{equation}
\operatorname{Bias}(\hat{\theta})
=
\frac{1}{R}\sum_{r=1}^{R}(\hat{\theta}_r-\theta_0),
\qquad
\operatorname{RMSE}(\hat{\theta})
=
\left\{
\frac{1}{R}\sum_{r=1}^{R}(\hat{\theta}_r-\theta_0)^2
\right\}^{1/2},
\end{equation}
where $\theta_0$ is the true parameter value. Smaller absolute bias and smaller RMSE indicate more accurate point estimation.

For interval estimation, we considered 95\% confidence intervals for the methods in the main-text comparison. Let $[L_r,U_r]$ denote the interval for $\theta$ from the $r$th successful replication. The coverage probability and mean interval length were defined as
\begin{equation}
\operatorname{CP}(\theta)
=
\frac{1}{R}\sum_{r=1}^{R} I\!\left(L_r \le \theta_0 \le U_r\right),
\qquad
\operatorname{MIL}(\theta)
=
\frac{1}{R}\sum_{r=1}^{R}(U_r-L_r).
\end{equation}
Coverage probability measures how often the interval contains the true value, whereas mean interval length reflects the precision of interval estimation.

To summarize interval performance more jointly, we also reported the interval score,
\begin{equation}
S_{\alpha}(L_r,U_r;\theta_0)
=
(U_r-L_r)
+
\frac{2}{\alpha}(L_r-\theta_0)I(\theta_0<L_r)
+
\frac{2}{\alpha}(\theta_0-U_r)I(\theta_0>U_r),
\end{equation}
where $\alpha=0.05$ for a 95\% interval. This measure combines interval width with a penalty for noncoverage, so smaller values indicate better interval performance. In addition, uncertainty calibration was assessed by the standard-error calibration ratio,
\begin{equation}
\operatorname{SECR}(\theta)
=
\frac{\operatorname{sd}(\hat{\theta}_1,\ldots,\hat{\theta}_R)}
{\frac{1}{R}\sum_{r=1}^{R}\widehat{\operatorname{SE}}_r(\theta)}.
\end{equation}
Values close to 1 indicate that the reported standard errors are well aligned with the observed Monte Carlo variability.

Variance-function recovery was evaluated by the mean squared error between the fitted variance and the true conditional variance across the design points,
\begin{equation}
\operatorname{Var\_MSE}
=
\frac{1}{R}\sum_{r=1}^{R}
\left\{
\frac{1}{n}\sum_{i=1}^{n}
\bigl(\hat{v}_r(s_i)-v(s_i)\bigr)^2
\right\},
\end{equation}
where $v(s_i)$ is the true conditional variance and $\hat{v}_r(s_i)$ is the fitted variance from replication $r$. Smaller values indicate closer agreement with the true variance function. Computational performance was summarized by the mean runtime in elapsed seconds over successful replications. Point-estimation results for the fixed-grid $n=50$ benchmark under Normal errors are reported in Table~\ref{tab:sim-point-normal}. The corresponding skewed-error robustness tables are deferred to Supplementary Table~\rev{\ref*{supp-tab:supp-main-point-skewed}} and Supplementary Table~\rev{\ref*{supp-tab:supp-main-diagnostic-skewed}}, while Supplementary Table~\rev{\ref*{supp-tab:supp-main-diagnostic-normal}} collects the full SECR and runtime summaries for the Normal-error benchmark.

\subsubsection{Results}\label{sec:single-sim-results}

The two main-text tables summarize point-estimation and interval-diagnostic performance for the fixed-grid $n=50$ benchmark under Normal errors. Supplementary Tables~\rev{\ref*{supp-tab:supp-main-diagnostic-skewed}--\ref*{supp-tab:supp-main-diagnostic-normal}} additionally collect the full extended diagnostics and the corresponding skewed-error robustness summaries. \rev{Denser-grid $n=100$ diagnostics are provided in Supplementary Tables~\ref*{supp-tab:supp-original-diagnostic-normal}--\ref*{supp-tab:supp-original-diagnostic-skewed}.} Across scenarios, standard NLS \rev{had larger RMSE and variance-function error than the heteroscedastic competitors in these generating scenarios}, and the \rev{profile} estimators remained competitive even when the working variance did not exactly match the data-generating form. \rev{All methods had small absolute biases; RMSE more clearly distinguished their estimation accuracy.}

Under Normal errors, all three \rev{profile} methods yielded smaller RMSE values than NLS. In the MM-type and Hill scenarios, the $S^{1/2}$ working variance gave the smallest RMSE values for both $V_{\max}$ and $K_m$. In the Exp scenario, the log-based working variance gave the smallest RMSE values for both parameters, with the $S^{1/3}$ model nearly tied for $V_{\max}$ and the $S^{1/2}$ model close for $K_m$. \rev{In the MM-type scenario, the best $V_{\max}$ RMSE decreased from 1.495 to 1.420, and the best $K_m$ RMSE from 0.946 to 0.886. Supplementary Section~\ref*{supp-sec:supp-matched} extends the comparison to weighted and transformation-based approaches.} The corresponding single-curve skewed-error robustness results, deferred to Supplementary Table~\rev{\ref*{supp-tab:supp-main-point-skewed}} and Supplementary Table~\rev{\ref*{supp-tab:supp-main-diagnostic-skewed}}, showed the same broad ranking.

\rev{The interval and diagnostic results distinguish interval width, coverage, and variance recovery. Coverage probabilities were usually within a few percentage points of the nominal 95\% level. For example, the square-root $K_m$ interval in the MM-type scenario shortened from 3.932 to 3.252, with coverage changing from 0.960 to 0.925. In the Hill scenario, coverage was 0.964 for that method and 0.972 for NLS.} Among the \rev{profile} methods, the log-based working variance produced the smallest Var\_MSE in the MM-type and Exp scenarios under both error distributions, whereas power-type working variances were more favorable in the Hill scenarios. \rev{Simple working shapes therefore improved variance recovery even when their forms differed from the generating variance; model choice can consider both variance recovery and interval calibration. For $R$ independent successful replications, the Monte Carlo standard error of coverage $\hat p$ is $\{\hat p(1-\hat p)/R\}^{1/2}$, approximately 0.007 near 0.95 when $R=1000$. Coverage differences should be read alongside this uncertainty and the additional paired comparisons.}

\begin{table}[H]
\centering
\caption{Point-estimation results under Normal errors across the MM-type, Exp, and Hill variance-generating scenarios for the fixed-grid $n=50$ benchmark. \rev{Bold entries indicate the smallest absolute bias and RMSE within each condition, including ties at the displayed precision.}}
\label{tab:sim-point-normal}
\scriptsize
\setlength{\tabcolsep}{4pt}
\MMtablefit{\linewidth}{%
\begin{tabular}{llrrrr}
\toprule
Scenario & Method & Bias($V_{\max}$) & RMSE($V_{\max}$) & Bias($K_m$) & RMSE($K_m$) \\
\midrule
MM-type & NLS & \textbf{0.00983} & 1.49511 & 0.00982 & 0.94643 \\
 & \rev{Profile:} $\log(S+1)$ & \rev{-0.01828} & 1.42498 & -0.01115 & 0.89041 \\
 & \rev{Profile:} $S^{1/2}$ & -0.03169 & \textbf{1.41982} & -0.01958 & \textbf{0.88589} \\
 & \rev{Profile:} $S^{1/3}$ & -0.01575 & 1.42392 & \textbf{-0.00885} & 0.88987 \\
\cmidrule(lr){1-6}
Exp & NLS & 0.00923 & 1.58807 & 0.01142 & 1.01002 \\
 & \rev{Profile:} $\log(S+1)$ & -0.00475 & \rev{\textbf{1.52924}} & \textbf{0.00083} & \textbf{0.95992} \\
 & \rev{Profile:} $S^{1/2}$ & -0.01123 & 1.53395 & -0.00336 & 0.96134 \\
 & \rev{Profile:} $S^{1/3}$ & \textbf{-0.00406} & 1.52959 & 0.00146 & \rev{0.96116} \\
\cmidrule(lr){1-6}
Hill & NLS & 0.02869 & 1.52565 & \textbf{0.00078} & 0.94184 \\
 & \rev{Profile:} $\log(S+1)$ & 0.01585 & 1.44450 & -0.00914 & 0.86827 \\
 & \rev{Profile:} $S^{1/2}$ & \textbf{0.00902} & \textbf{1.42271} & -0.01380 & \textbf{0.84891} \\
 & \rev{Profile:} $S^{1/3}$ & \rev{0.01567} & 1.44295 & -0.00912 & 0.86886 \\
\bottomrule
\end{tabular}%
}
\end{table}
\FloatBarrier

\clearpage
\begin{landscape}
\thispagestyle{plain}
\begingroup
\MMtablefont
\setlength{\tabcolsep}{5pt}
\refstepcounter{table}\label{tab:sim-diagnostic-normal}
\begin{center}
\parbox{0.97\linewidth}{\MMtablecaptionfont\centering\textbf{Table \thetable}\\Interval and variance diagnostics under Normal errors across the MM-type, Exp, and Hill variance-generating scenarios for the fixed-grid $n=50$ benchmark. CP = coverage probability; MIL = mean interval length; IS = interval score. Full SECR and runtime summaries are reported in Supplementary Table~\rev{\ref*{supp-tab:supp-main-diagnostic-normal}}. Bold entries indicate the smallest interval score and variance MSE among the main-text methods for each scenario and parameter.}

\vspace{0.4em}
\MMtablefit{0.97\linewidth}{%
\begin{tabular}{llccccccc}
\toprule
Scenario & Method & CP($V_{\max}$) & MIL($V_{\max}$) & IS($V_{\max}$) & CP($K_m$) & MIL($K_m$) & IS($K_m$) & Var\_MSE \\
\midrule
MM-type & NLS & 0.94300 & 5.72523 & 7.04436 & 0.96000 & 3.93150 & 4.42846 & 5.12201 \\
 & \rev{Profile:} $\log(S+1)$ & 0.93400 & 5.43209 & \rev{6.75408} & 0.94100 & 3.45392 & 4.17841 & \textbf{2.03549} \\
 & \rev{Profile:} $S^{1/2}$ & 0.93900 & \rev{5.44576} & \textbf{6.70244} & 0.92500 & \rev{3.25181} & 4.19760 & 2.97455 \\
 & \rev{Profile:} $S^{1/3}$ & 0.94000 & 5.45422 & 6.72346 & 0.94100 & 3.42665 & \rev{\textbf{4.16862}} & 2.09529 \\
\cmidrule(lr){1-9}
Exp & NLS & 0.94900 & 6.25973 & 7.56320 & 0.96500 & 4.29880 & 4.93263 & 7.91152 \\
 & \rev{Profile:} $\log(S+1)$ & 0.94800 & \rev{5.93343} & \rev{\textbf{7.34822}} & 0.94300 & 3.77361 & \textbf{4.68422} & \textbf{3.24468} \\
 & \rev{Profile:} $S^{1/2}$ & 0.93800 & \rev{5.95002} & \rev{7.35021} & 0.93100 & \rev{3.55414} & \rev{4.72940} & 4.57808 \\
 & \rev{Profile:} $S^{1/3}$ & 0.94600 & 5.96452 & 7.36078 & 0.94300 & 3.74806 & 4.70215 & 3.42358 \\
\cmidrule(lr){1-9}
Hill & NLS & 0.94300 & 5.99868 & 7.04804 & 0.97200 & 4.11734 & 4.38003 & 8.95651 \\
 & \rev{Profile:} $\log(S+1)$ & 0.94300 & 5.60929 & \rev{6.65360} & 0.97100 & 3.56541 & 3.94124 & 3.48907 \\
 & \rev{Profile:} $S^{1/2}$ & 0.95300 & \rev{5.56969} & \textbf{6.60195} & 0.96400 & 3.32490 & \textbf{3.80043} & \textbf{3.24069} \\
 & \rev{Profile:} $S^{1/3}$ & 0.95000 & 5.63719 & \rev{6.64772} & 0.96900 & 3.54028 & 3.92400 & 3.42797 \\
\bottomrule
\end{tabular}
}
\end{center}
\endgroup
\end{landscape}
\clearpage
\FloatBarrier

\input{sections/additional_simulation_rev1}

\subsection{Clustered simulation benchmark}\label{sec:cluster-simulation}

\subsubsection{Design and competing estimators}\label{sec:cluster-sim-design}

\rev{All methods were assessed using the same checks, and only converged fits with valid kinetic-parameter estimates, standard errors and variance estimates were included. Software versions, detailed inclusion criteria, random seeds and counts of included and excluded runs are documented in Supplementary Section~\ref*{supp-sec:supp-clustered} and the journal-submission reproduction archive.}

To evaluate the repeated-measurement formulation on comparable footing with the single-curve benchmark, we conducted a second simulation calibrated to the empirical assay layout of the Alves AP data. Two clustered templates were considered: a harsher $m=3$ design corresponding to one depth slice and a more stable $m=6$ design corresponding to two pooled depth slices. In both templates, each cluster was observed at eight substrate concentrations $S\in\{10,40,80,130,200,350,700,1200\}$ with four technical replicates per concentration. \rev{The nominal generating values were $(V_{\max},K_m)=(100,20)$ (height floor: Supplementary Section~\ref*{supp-sec:supp-clustered})}; cluster-to-cluster heterogeneity entered through a random effect on $V_{\max}$ calibrated from the corresponding Alves template, and the within-cluster residual variance followed the heteroscedastic working form $\gamma h(S)$ with $h(S)=S^{1/2}$. Each setting was replicated 1,000 times.

We compared five estimators. The first two were pooled analyses that ignored clustering: a homoscedastic \rev{NLS-equivalent} fit and a pooled heteroscedastic fit with $h(S)=S^{1/2}$. \rev{Both pooled rows use \texttt{fit\_mm()}, with \texttt{variance="constant"} or \texttt{"sqrt"}, respectively.} The third was the \rev{working profile/moment} clustered \rev{fit in Section~\ref{sec:cluster-algorithm}, with the random effect and covariance specified in \eqref{eq:cluster-model}--\eqref{eq:cluster-covariance}}. The last two comparators were fully parametric NLME fits with a random intercept on $V_{\max}$, one with constant residual variance and one with a \texttt{varConstPower} residual structure. To keep the clustered benchmark readable in the main text, Tables~\ref{tab:clustered-point}--\ref{tab:clustered-diagnostic} focus on the matched Normal/Normal setting for the latent random effects and residual errors. The corresponding matched Skewed/Skewed robustness tables, together with the exact skewed data-generating details, are deferred to Supplementary Tables~\rev{\ref*{supp-tab:supp-clustered-point-skewed}--\ref*{supp-tab:supp-clustered-diagnostic-skewed}} and the Supplementary Material discussion. \rev{This simulation evaluated the performance of the standard errors and 95\% confidence intervals calculated by the default clustered profile method in Section~\ref{sec:cluster-algorithm}.}

\enlargethispage{\baselineskip}
\begin{table}[H]
\centering
\caption{Point-estimation results for the clustered benchmark under the matched Normal/Normal latent-error setting. The two template designs are calibrated from the Alves AP assay layout with either 3 or 6 clusters, 8 substrate levels, and 4 technical replicates per level. \rev{Bold entries indicate the smallest RMSE within each condition, including ties at the displayed precision.}}
\label{tab:clustered-point}
\scriptsize
\setlength{\tabcolsep}{4pt}
\MMtablefit{\linewidth}{%
\begin{tabular}{llrr}
\toprule
Scenario & Method & RMSE($V_{\max}$) & RMSE($K_m$) \\
\midrule
3 clusters & Pooled NLS & 24.038 & 2.235 \\
 & Pooled $S^{1/2}$ & \rev{\textbf{23.952}} & 1.761 \\
 & Clustered $S^{1/2}$ & \rev{23.963} & \rev{1.622} \\
 & NLME (homoscedastic) & 24.039 & 2.060 \\
 & NLME (const+power) & \rev{23.962} & \rev{\textbf{1.621}} \\
\cmidrule(lr){1-4}
6 clusters & Pooled NLS & \rev{\textbf{19.796}} & 1.381 \\
 & Pooled $S^{1/2}$ & 19.810 & 1.047 \\
 & Clustered $S^{1/2}$ & \rev{\textbf{19.796}} & \rev{\textbf{0.971}} \\
 & NLME (homoscedastic) & 19.801 & 1.222 \\
 & NLME (const+power) & \rev{19.799} & 0.991 \\
\bottomrule
\end{tabular}%
}
\end{table}

\begin{table}[H]
\centering
\caption{Coverage and variance-recovery results for the clustered benchmark under the matched Normal/Normal latent-error setting. CP = empirical 95\% Wald coverage probability. Var\_RMSE denotes the RMSE of the fitted marginal variance surface across the design points. RMSE($\tau^2$) is reported only for methods that explicitly estimate a random-effect variance. \rev{Bold entries indicate the smallest RMSE within each condition, including ties at the displayed precision.}}
\label{tab:clustered-diagnostic}
\scriptsize
\setlength{\tabcolsep}{4pt}
\MMtablefit{\linewidth}{%
\begin{tabular}{llcccc}
\toprule
Scenario & Method & CP($V_{\max}$) & CP($K_m$) & Var\_RMSE & RMSE($\tau^2$) \\
\midrule
3 clusters & Pooled NLS & 0.312 & 0.995 & 1045.4 & --- \\
 & Pooled $S^{1/2}$ & 0.343 & 0.989 & 1074.0 & --- \\
 & Clustered $S^{1/2}$ & \rev{0.766} & \rev{0.932} & \rev{880.6} & \rev{\textbf{1392.0}} \\
 & NLME (homoscedastic) & 0.765 & 0.975 & 901.1 & 1394.2 \\
 & NLME (const+power) & \rev{0.762} & 0.930 & \rev{\textbf{877.7}} & \rev{1392.5} \\
\cmidrule(lr){1-6}
6 clusters & Pooled NLS & 0.339 & 1.000 & 1006.2 & --- \\
 & Pooled $S^{1/2}$ & 0.362 & 1.000 & 1009.9 & --- \\
 & Clustered $S^{1/2}$ & 0.855 & 0.953 & 729.5 & 1159.7 \\
 & NLME (homoscedastic) & 0.857 & 0.985 & 746.4 & \rev{\textbf{1158.4}} \\
 & NLME (const+power) & 0.853 & 0.943 & \rev{\textbf{728.1}} & \rev{1159.6} \\
\bottomrule
\end{tabular}%
}
\end{table}

\subsubsection{Results}\label{sec:cluster-sim-results}

\rev{Tables~\ref{tab:clustered-point}--\ref{tab:clustered-diagnostic} show that accounting for clustering primarily changes uncertainty estimates: the point-estimation RMSE for $V_{\max}$ is similar across methods, while its coverage increases from about 0.31--0.36 for pooled fits to about 0.76 with three clusters and 0.85 with six clusters for the working profile/moment and NLME fits. These coverages remain below the nominal 95\% level, identifying small-cluster calibration as a consideration for both approaches. For the clustered working fit, coverage 0.855 with six clusters has Monte Carlo standard error about 0.011. In this design the random effect acts directly on $V_{\max}$, and pooled $K_m$ intervals are overconservative; other random-effect specifications may affect the two parameters differently.}

\rev{The cluster-aware procedures also estimated the reference marginal variance curve more accurately than the pooled alternatives (Supplementary Section~\ref*{supp-sec:supp-clustered}).} In the 6-cluster setting, for example, the marginal variance-surface RMSE was about 1006--1010 for the pooled methods but only about 728--746 for the clustered and NLME fits. The matched Skewed/Skewed robustness tables in Supplementary Tables~\rev{\ref*{supp-tab:supp-clustered-point-skewed}--\ref*{supp-tab:supp-clustered-diagnostic-skewed}} show the same qualitative ordering. \rev{The practical benefit is a clearer separation of cluster-level variation from concentration-dependent residual noise, alongside stable mean-curve estimation.}

Within the cluster-aware group, the \rev{working profile/moment} estimator and the NLME benchmarks were broadly comparable for $V_{\max}$, but the \rev{working} clustered fit was often slightly more competitive for $K_m$ RMSE than the homoscedastic NLME fit and remained close to the more flexible \texttt{varConstPower} alternative. \rev{For example, the six-cluster $K_m$ RMSE values were 0.971 and 0.991 for the working fit and the flexible NLME fit, respectively. These results support the low-dimensional MM formulation as an alternative with comparable estimation accuracy under the studied designs.}

\section{Real enzyme data applications}\label{sec:real-data}

We apply the proposed framework to two publicly available enzyme-kinetic data sets. In each application we fit the same four candidate models---three \rev{Gaussian-working} heteroscedastic variants and standard homoscedastic NLS---and compare them by \rev{working} AIC, BIC, fitted curves, and the resulting estimates of $V_{\max}$ and $K_m$. \rev{Heteroscedastic models were fitted using \texttt{inferMM} 0.1.2 and homoscedastic models using \texttt{stats::nls()}. Model-conditional Wald intervals used standard-normal critical values.}

\subsection{Dataset 1: Self-Driving Laboratories (SDL) -- Protein Fitness Landscape}

\subsubsection{Data description and preprocessing}

The first data set comes from the SAMPLE platform of Rapp et al.~\cite{rapp2024}. We analyze the six parental GH1 sequences (variants 1111--6666) for which Michaelis--Menten kinetic data were reported. Initial reaction velocities were estimated from the first three fluorescence time points for each replicate \rev{and converted to $\mu$M/min using the published fluorescence calibration curve}. This yields replicate-level $(S,v)$ pairs at \rev{eight} substrate concentrations per variant, with \rev{four} analytical replicates per concentration. \rev{Specifically, $S\in\{3.90625,7.8125,15.625,31.25,62.5,125,250,500\}$ $\mu$M, giving 32 observations per variant. All initial linear-fit $R^2$ values in the processed panel exceed 0.97.}

\subsubsection{Empirical assessment of the variance structure}

Plots of replicate-level velocities against $S$ showed a consistent increase in spread with substrate concentration. In the SDL data this pattern is moderate and broadly consistent with a power-type variance structure, which is in line with classical weighting and transformation approaches for enzyme-kinetic data under nonconstant variance \cite{askelof1976,ruppert1989} and motivates the monotone working variance models considered below.

\subsubsection{Model fitting and comparison}\label{sec:sdl-fitting}

\rev{We fitted \eqref{eq:mm-mean} separately for each enzyme variant using \texttt{inferMM} 0.1.2. Reaction-level initial-rate errors were assumed conditionally independent within each variant. Intervals are conditional on this assumption and the working variance model, without a separate component for shared experimental or between-preparation biological variation.} We compared four prespecified variance forms: standard NLS, $h(S)=S^{1/2}$, $h(S)=S^{1/3}$, and $h(S)=\log(S+1)$. \rev{Working AIC and BIC rank these candidates; optional screening over power-family forms $h(S)=S^p$ is also available. The \texttt{group\_mm()} interface organizes separate \texttt{fit\_mm()} analyses rather than modelling within-cluster dependence; \texttt{stats::nls()} provides the constant-variance comparator.}

\begin{figure}[H]
\centering
\includegraphics[width=0.96\textwidth]{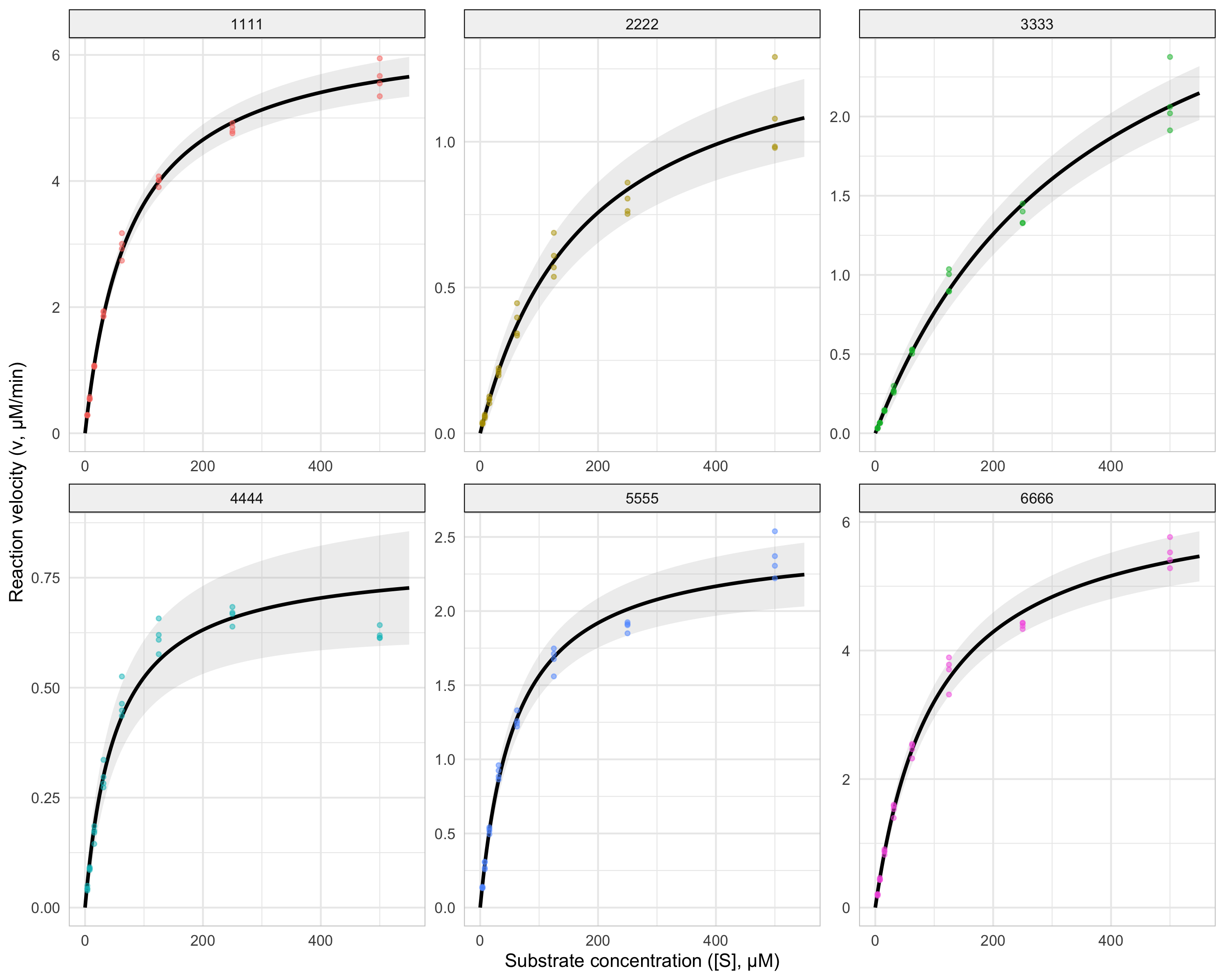}
\caption{Fitted Michaelis--Menten curves for the best-fitting model [$S^{1/2}$] across the six SDL enzyme variants (1111--6666). The black line shows the fitted mean curve; the shaded region represents the \rev{Gaussian-working residual envelope conditional on the fitted parameters, clipped below at zero}. Points indicate individual replicate observations\rev{, with 32 observations at eight concentrations per variant}. \rev{Velocities are in $\mu$M/min.}}
\label{fig:sdl-best}
\end{figure}

\rev{Table~\ref{tab:sdl-model-comparison} reports all six variants. Each heteroscedastic form had lower working AIC and BIC than NLS, with the square-root form lowest throughout. Figure~\ref{fig:sdl-best} and Supplementary Figure~\ref*{supp-fig:supp-sdl-nls} contrast the fits and residual envelopes. The envelope $\hat\mu(s)\pm1.96\sqrt{\hat\gamma h(s)}$ shows residual variability conditional on fitted parameters, with its lower bound clipped at zero.}

\begin{landscape}
\begin{table}[H]
\centering
\caption{\rev{SDL estimates and model comparison for six GH1 variants (32 observations at eight concentrations each). Bold rows have the lowest AIC within each variant; Rank follows within-variant AIC, and rows use a fixed model order. CI = 95\% confidence interval; $\hat\gamma$ = estimated dispersion; NLS = homoscedastic nonlinear least squares.}}
\label{tab:sdl-model-comparison}
\renewcommand{\MMtablerowstretch}{1.05}
\tiny
\setlength{\tabcolsep}{4pt}
\MMtablefit{\linewidth}{%
\begin{tabular}{llrrrrrrrc}
\toprule
Enzyme & Model & $\hat V_{\max}$ & 95\% CI ($V_{\max}$) & $\hat K_m$ & 95\% CI ($K_m$) & $\hat\gamma$ & AIC & BIC & Rank \\
\midrule
\textbf{1111} & NLS & \rev{6.417} & \rev{[6.267, 6.567]} & \rev{75.85} & \rev{[70.59, 81.12]} & \rev{0.01262} & \rev{-43.10} & \rev{-38.70} & 4 \\
 & $\log(S+1)$ & \rev{6.429} & \rev{[6.280, 6.578]} & \rev{76.42} & \rev{[71.79, 81.05]} & \rev{0.00256} & \rev{-53.98} & \rev{-49.59} & 3 \\
 & \textbf{$S^{1/2}$} & \rev{\textbf{6.445}} & \rev{\textbf{[6.282, 6.609]}} & \rev{\textbf{76.96}} & \rev{\textbf{[72.60, 81.32]}} & \rev{\textbf{0.00110}} & \rev{\textbf{-60.71}} & \rev{\textbf{-56.31}} & \textbf{1} \\
 & $S^{1/3}$ & \rev{6.428} & \rev{[6.274, 6.582]} & \rev{76.37} & \rev{[71.87, 80.86]} & \rev{0.00237} & \rev{-56.26} & \rev{-51.86} & 2 \\
\cmidrule(lr){1-10}
\textbf{2222} & NLS & 1.459 & \rev{[1.313, 1.605]} & \rev{185.97} & \rev{[143.48, 228.47]} & \rev{0.003212} & \rev{-86.89} & \rev{-82.50} & 4 \\
 & $\log(S+1)$ & \rev{1.448} & \rev{[1.314, 1.582]} & \rev{182.51} & \rev{[146.39, 218.63]} & \rev{0.000572} & \rev{-101.89} & \rev{-97.49} & 3 \\
 & \textbf{$S^{1/2}$} & \rev{\textbf{1.433}} & \rev{\textbf{[1.304, 1.563]}} & \rev{\textbf{178.64}} & \rev{\textbf{[147.52, 209.76]}} & \rev{\textbf{0.000198}} & \rev{\textbf{-115.49}} & \rev{\textbf{-111.10}} & \textbf{1} \\
 & $S^{1/3}$ & \rev{1.441} & \rev{[1.309, 1.572]} & \rev{180.60} & \rev{[146.69, 214.50]} & \rev{0.000491} & \rev{-106.59} & \rev{-102.19} & 2 \\
\cmidrule(lr){1-10}
\textbf{3333} & NLS & \rev{3.650} & \rev{[3.235, 4.065]} & \rev{381.55} & \rev{[303.13, 459.98]} & \rev{0.005682} & \rev{-68.64} & \rev{-64.24} & 4 \\
 & $\log(S+1)$ & \rev{3.631} & \rev{[3.254, 4.007]} & \rev{377.72} & \rev{[309.21, 446.23]} & \rev{0.000984} & \rev{-84.53} & \rev{-80.13} & 3 \\
 & \textbf{$S^{1/2}$} & \rev{\textbf{3.605}} & \rev{\textbf{[3.253, 3.958]}} & \rev{\textbf{373.11}} & \rev{\textbf{[313.20, 433.01]}} & \rev{\textbf{0.000322}} & \rev{\textbf{-99.93}} & \rev{\textbf{-95.53}} & \textbf{1} \\
 & $S^{1/3}$ & \rev{3.613} & \rev{[3.248, 3.977]} & \rev{374.37} & \rev{[310.04, 438.71]} & \rev{0.000826} & \rev{-89.93} & \rev{-85.54} & 2 \\
\cmidrule(lr){1-10}
\textbf{4444} & NLS & \rev{0.746} & \rev{[0.703, 0.789]} & \rev{42.52} & \rev{[34.14, 50.89]} & \rev{0.001855} & \rev{-104.46} & \rev{-100.06} & 4 \\
 & $\log(S+1)$ & \rev{0.766} & \rev{[0.720, 0.813]} & \rev{47.17} & \rev{[39.03, 55.30]} & \rev{0.000417} & \rev{-112.01} & \rev{-107.61} & 3 \\
 & \textbf{$S^{1/2}$} & \rev{\textbf{0.795}} & \rev{\textbf{[0.742, 0.849]}} & \rev{\textbf{52.00}} & \rev{\textbf{[43.73, 60.27]}} & \rev{\textbf{0.000184}} & \rev{\textbf{-117.74}} & \rev{\textbf{-113.34}} & \textbf{1} \\
 & $S^{1/3}$ & \rev{0.777} & \rev{[0.728, 0.826]} & \rev{48.74} & \rev{[40.56, 56.91]} & \rev{0.000393} & \rev{-113.70} & \rev{-109.30} & 2 \\
\cmidrule(lr){1-10}
\textbf{5555} & NLS & \rev{2.532} & \rev{[2.432, 2.632]} & \rev{62.24} & \rev{[54.55, 69.92]} & \rev{0.006995} & \rev{-61.99} & \rev{-57.59} & 4 \\
 & $\log(S+1)$ & \rev{2.513} & \rev{[2.418, 2.608]} & \rev{60.69} & \rev{[54.47, 66.91]} & \rev{0.001334} & \rev{-74.79} & \rev{-70.39} & 3 \\
 & \textbf{$S^{1/2}$} & \rev{\textbf{2.488}} & \rev{\textbf{[2.392, 2.584]}} & \rev{\textbf{59.18}} & \rev{\textbf{[53.88, 64.47]}} & \rev{\textbf{0.000514}} & \rev{\textbf{-84.92}} & \rev{\textbf{-80.52}} & \textbf{1} \\
 & $S^{1/3}$ & \rev{2.499} & \rev{[2.404, 2.593]} & \rev{59.77} & \rev{[53.96, 65.59]} & \rev{0.001185} & \rev{-78.40} & \rev{-74.00} & 2 \\
\cmidrule(lr){1-10}
\textbf{6666} & NLS & \rev{6.498} & \rev{[6.265, 6.730]} & \rev{102.72} & \rev{[92.68, 112.76]} & \rev{0.020682} & \rev{-27.30} & \rev{-22.90} & 4 \\
 & $\log(S+1)$ & \rev{6.491} & \rev{[6.263, 6.719]} & \rev{102.48} & \rev{[93.64, 111.32]} & \rev{0.004099} & \rev{-38.87} & \rev{-34.47} & 3 \\
 & \textbf{$S^{1/2}$} & \rev{\textbf{6.481}} & \rev{\textbf{[6.237, 6.725]}} & \rev{\textbf{102.18}} & \rev{\textbf{[93.96, 110.41]}} & \rev{\textbf{0.001687}} & \rev{\textbf{-46.88}} & \rev{\textbf{-42.48}} & \textbf{1} \\
 & $S^{1/3}$ & \rev{6.479} & \rev{[6.244, 6.713]} & \rev{102.00} & \rev{[93.43, 110.57]} & \rev{0.003775} & \rev{-41.32} & \rev{-36.92} & 2 \\
\bottomrule
\end{tabular}%
}
\end{table}
\end{landscape}
\FloatBarrier

\subsubsection{Interpretation of fitted results}\label{sec:sdl-interpretation}

\rev{The estimated kinetic parameters distinguish the six GH1 variants, with similar point estimates under the square-root model and NLS.} On average, NLS produces slightly larger estimates of both $V_{\max}$ and $K_m$, consistent with equal weighting giving somewhat greater influence to high-concentration observations where variability is higher. \rev{The square-root model generally gives narrower model-conditional confidence intervals for $K_m$ than NLS, with similar widths for $V_{\max}$. We compare fitted variances $\hat\gamma h(s)$ at common substrate concentrations.}

\subsection{Dataset 2: Alves et al. (2021) -- Soil exoenzyme kinetics}

\subsubsection{Data description and preprocessing}

The second data set comes from the Supplementary Material accompanying Alves et al.~\cite{alves2021}. It contains Michaelis--Menten measurements for three microbial exoenzymes---acid phosphatase (AP), beta-glucosidase (BG), and leucine aminopeptidase (LAP)---across multiple soil depths, temperatures, biological cores, and substrate concentrations. After removing the metadata header row, coercing numeric fields, and excluding sentinel missing values ($-9999$), we pool observations across depth and temperature within each enzyme to obtain a descriptive variance-aware comparison on the concentration--response scale. \rev{The resulting parameters summarize the observed assay conditions. The pooled variation combines concentration-dependent heteroscedasticity with between-condition differences in the mean and variance, allowing a descriptive assessment of variance-aware fitting across heterogeneous conditions. Condition-specific biochemical interpretation would require a hierarchical analysis retaining those factors.}

\subsubsection{Empirical assessment of the variance structure}

The Alves data show much stronger heteroscedasticity than the SDL data: replicate activity measurements spread out rapidly as substrate concentration increases. Alves et al.~\cite{alves2021} addressed this issue with a Box--Cox transformation, whereas here we model it directly through monotone working variance functions.

\subsubsection{Model fitting and comparison}

\begin{figure}[H]
\centering
\includegraphics[width=0.96\textwidth]{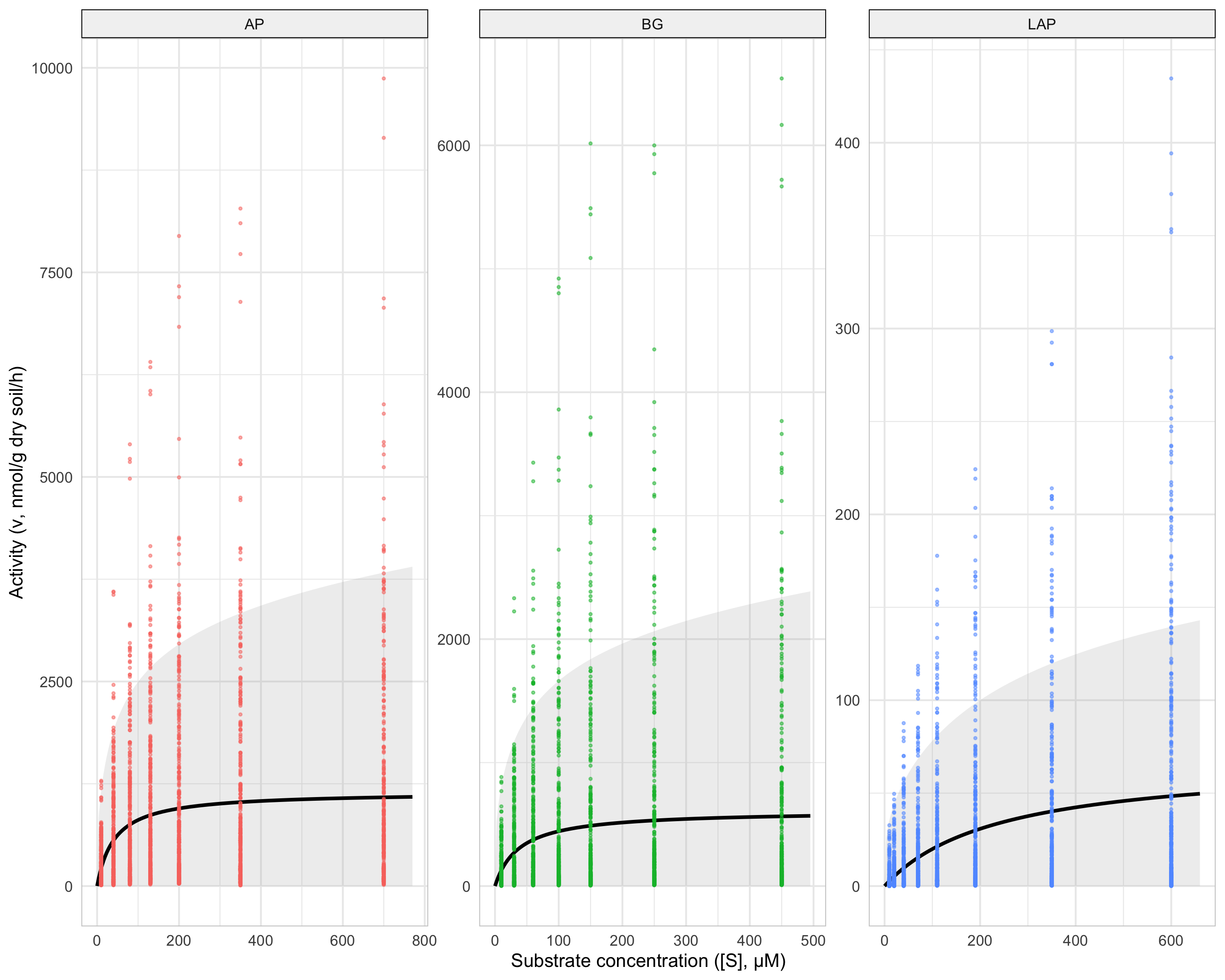}
\caption{Fitted Michaelis--Menten curves for the best-fitting model [$S^{1/2}$] for AP, BG, and LAP in the Alves et al. (2021) data set. The shaded band expands with increasing substrate concentration, proportionally reflecting the heteroscedastic variance structure. \rev{The residual envelope is conditional on the fitted parameters and clipped at zero.} Points indicate pooled replicate observations across depth layers and temperatures.}
\label{fig:alves-best}
\end{figure}

The same four-model comparison was applied to the pooled Alves data using \rev{the supplied scripts and \texttt{inferMM} 0.1.2, separately within each enzyme}. \rev{The grouped workflow corresponds to that organized by \texttt{group\_mm()}; Table~\ref{tab:alves-model-comparison} reports the pooled analysis. It summarizes concentration--response relationships across depth and temperature, with core-level dependence addressed separately in the clustered formulation.} The square-root variance model again achieved the lowest \rev{working} AIC and BIC for all three enzymes, and the \rev{criterion difference from} NLS was much larger than in the SDL analysis. \rev{Figure~\ref{fig:alves-best} and Supplementary Figure~\ref*{supp-fig:supp-alves-nls} contrast the fitted curves and residual envelopes. The constant-variance NLS envelope is wide because one residual scale absorbs extreme high-concentration variation. The criteria assess working-model fit across concentrations and environmental conditions.}

\begin{landscape}
\begin{table}[H]
\centering
\caption{Alves et al. (2021) data set: parameter estimates and model comparison statistics for three soil exoenzymes (AP, BG, LAP). Bold rows indicate the best-fitting model per enzyme according to AIC. Rows are displayed in a fixed model order, and Rank is based on within-enzyme AIC. $V_{\max}$ units: nmol/g dry soil/h; $K_m$ units: $\mu$M; NLS = standard homoscedastic nonlinear least squares.}
\label{tab:alves-model-comparison}
\scriptsize
\setlength{\tabcolsep}{4pt}
\MMtablefit{\linewidth}{%
\begin{tabular}{llrrrrrrrc}
\toprule
Enzyme & Model & $\hat V_{\max}$ & 95\% CI ($V_{\max}$) & $\hat K_m$ & 95\% CI ($K_m$) & $\hat\gamma$ & AIC & BIC & Rank \\
\midrule
\textbf{AP} & NLS & 1181.7 & [1087.7, 1275.7] & 47.40 & [32.39, 62.42] & 1044213 & 47091 & 47108 & 4 \\
 & $\log(S+1)$ & 1168.8 & [1077.3, 1260.2] & 45.04 & \rev{[32.49, 57.59]} & 195566 & 46604 & 46622 & 3 \\
 & \textbf{$S^{1/2}$} & \textbf{1149.8} & \textbf{[1058.2, 1241.4]} & \textbf{42.51} & \textbf{[32.11, 52.91]} & \textbf{74230} & \textbf{46224} & \textbf{46242} & \textbf{1} \\
 & $S^{1/3}$ & 1163.5 & \rev{[1070.8, 1256.1]} & 44.41 & \rev{[32.55, 56.27]} & 174942 & 46445 & 46463 & 2 \\
\cmidrule(lr){1-10}
\textbf{BG} & NLS & 601.8 & [537.4, 666.3] & 35.83 & [20.82, 50.84] & 479871 & 44245 & 44263 & 4 \\
 & $\log(S+1)$ & 606.2 & \rev{[541.8, 670.6]} & 36.88 & \rev{[23.59, 50.17]} & 94591 & 43768 & 43785 & 3 \\
 & \textbf{$S^{1/2}$} & \textbf{612.2} & \textbf{[545.3, 679.1]} & \textbf{38.00} & \rev{\textbf{[25.98, 50.03]}} & \textbf{38685} & \textbf{43395} & \textbf{43413} & \textbf{1} \\
 & $S^{1/3}$ & 609.3 & \rev{[543.4, 675.2]} & 37.46 & \rev{[24.52, 50.40]} & 87909 & 43627 & 43645 & 2 \\
\cmidrule(lr){1-10}
\textbf{LAP} & NLS & 73.02 & \rev{[61.90, 84.14]} & \rev{282.2} & \rev{[193.2, 371.3]} & 1457.0 & 32741 & 32759 & 4 \\
 & $\log(S+1)$ & 70.88 & [60.69, 81.08] & 263.9 & [187.8, 340.0] & 257.6 & 31814 & 31833 & 3 \\
 & \textbf{$S^{1/2}$} & \textbf{67.74} & \textbf{[58.32, 77.17]} & \textbf{240.1} & \textbf{[177.8, 302.5]} & \textbf{88.32} & \textbf{30828} & \textbf{30846} & \textbf{1} \\
 & $S^{1/3}$ & 69.60 & \rev{[59.74, 79.46]} & 254.1 & [184.2, 324.1] & 220.1 & 31396 & 31415 & 2 \\
\bottomrule
\end{tabular}%
}
\end{table}
\end{landscape}
\FloatBarrier

\subsubsection{Interpretation of fitted results}\label{sec:alves-interpretation}

The pooled AP, BG, and LAP curves differ markedly in both scale and apparent affinity, but the central message here is comparative rather than biochemical. Relative to the square-root model, NLS \rev{gives larger $V_{\max}$ and $K_m$ estimates for AP and LAP and smaller estimates for BG}. \rev{Variance comparisons use the fitted variance $\hat\gamma h(s)$, residual patterns, and working likelihood, accounting for the different normalizations of $h$. Model-conditional interval changes are more apparent for $K_m$ than for $V_{\max}$. This descriptive comparison shows improved working criteria for variance-aware fitting to the pooled concentration--response data, whose variability includes both concentration and environmental effects.}

\subsection{Comparative summary}

Table~\ref{tab:cross-dataset-summary} summarizes mean AIC, mean BIC, and average parameter estimates across both data sets. The square-root variance model consistently provided the best fit by \rev{working} AIC and BIC \rev{among the four candidates}, and the ordering $S^{1/2} \succ S^{1/3} \succ \log(S+1) \succ \mathrm{NLS}$ was preserved across all nine enzyme--data-set combinations.

\begin{table}[H]
\centering
\caption{Cross-data-set summary of mean AIC, mean BIC, and average estimated parameters. Bold rows indicate the best-fitting model per data set according to mean AIC and mean BIC.}
\label{tab:cross-dataset-summary}
\scriptsize
\setlength{\tabcolsep}{4pt}
\MMtablefit{\linewidth}{%
\begin{tabular}{llrrrr}
\toprule
Dataset & Model & Mean AIC & Mean BIC & Avg. $\hat V_{\max}$ & Avg. $\hat K_m$ \\
\midrule
SDL (n=6 enzymes) & NLS & \rev{-65.40} & \rev{-61.00} & \rev{3.550} & \rev{141.8} \\
 & $\log(S+1)$ & \rev{-77.68} & \rev{-73.28} & \rev{3.546} & \rev{141.2} \\
 & \textbf{$S^{1/2}$} & \rev{\textbf{-87.61}} & \rev{\textbf{-83.21}} & \rev{\textbf{3.541}} & \rev{\textbf{140.3}} \\
 & $S^{1/3}$ & \rev{-81.03} & \rev{-76.64} & \rev{3.539} & \rev{140.3} \\
\cmidrule(lr){1-6}
Alves (n=3 enzymes) & NLS & 41359 & 41377 & 618.8 & 121.8 \\
 & $\log(S+1)$ & 40729 & 40747 & 615.3 & 115.3 \\
 & \textbf{$S^{1/2}$} & \textbf{40149} & \textbf{40167} & \textbf{609.9} & \textbf{106.9} \\
 & $S^{1/3}$ & 40489 & 40507 & 614.1 & 112.0 \\
\bottomrule
\end{tabular}%
}
\end{table}
\FloatBarrier

The contrast is much larger in the Alves data than in the SDL data, which is consistent with the \rev{greater observed heterogeneity} in the soil assays. \rev{On average, NLS gives higher $V_{\max}$ and $K_m$ estimates than the best heteroscedastic fit, with the direction varying among individual curves. Overall, the real-data results favor the square-root shape among the prespecified candidates and illustrate how variance modeling informs both model comparison and kinetic-parameter inference. The pooled Alves estimates summarize the observed depth and temperature conditions within each enzyme.}

\input{sections/discussion_rev1}

\section*{\rev{Funding}}

\rev{This research received no external funding.}

\Needspace{6\baselineskip}
\section*{Declaration of competing interest}

The authors declare that they have no known competing financial interests or personal relationships that could have appeared to influence the work reported in this paper.

\section*{Declaration of the use of generative AI and AI-assisted technologies}

\rev{The authors used OpenAI Codex for manuscript editing and assistance with analysis code. The authors take full responsibility for the analyses and final manuscript.}

\section*{\rev{Related manuscript}}

\rev{A related manuscript, ``inferMM: An R Package for Variance-Aware
Michaelis--Menten Estimation and Inference,'' has been submitted to
\textit{The R Journal}. It focuses on the software interface and analysis
workflows and shares the Michaelis--Menten framework and illustrative
datasets with the present article. The present article focuses on the
statistical formulation and inferential evaluation, including the
higher-moment extension.}

\section*{Data availability}

The raw SDL data analyzed in this study are publicly available from the Zenodo repository associated with Rapp et al.~\cite{rapp2024} (\url{https://doi.org/10.5281/zenodo.10048592}). The soil exoenzyme data are available in the Supplementary Material accompanying Alves et al.~\cite{alves2021}. The R package \texttt{inferMM} is available on CRAN at \url{https://CRAN.R-project.org/package=inferMM}. The supplementary material and \textit{Data and Software} reproduction archive referred to in this article were prepared for the journal submission and are not included in this preprint.

\Needspace{6\baselineskip}
\bibliographystyle{elsarticle-num}
\begingroup
\let\originalbibitem\bibitem
\renewcommand{\bibitem}[1]{%
  \ifshowrevisions
    \color{black}%
    \ifstrequal{#1}{wu2025}{\color{blue}}{}%
    \ifstrequal{#1}{neun2022}{\color{blue}}{}%
    \ifstrequal{#1}{marcel1998}{\color{blue}}{}%
    \ifstrequal{#1}{ritz2015}{\color{blue}}{}%
    \ifstrequal{#1}{carroll1988}{\color{blue}}{}%
    \ifstrequal{#1}{kim2019}{\color{blue}}{}%
    \ifstrequal{#1}{tsiatis2006}{\color{blue}}{}%
    \ifstrequal{#1}{davidian1995}{\color{blue}}{}%
    \ifstrequal{#1}{golub1973}{\color{blue}}{}%
    \ifstrequal{#1}{zeileis2006}{\color{blue}}{}%
    \ifstrequal{#1}{liang1986}{\color{blue}}{}%
  \fi
  \originalbibitem{#1}}
\bibliography{references}
\endgroup

\input{sections/figure_legends_rev1}

\end{document}

%% file: sections/table_layout_rev1.tex
\usepackage{adjustbox}
\usepackage{caption}
\usepackage{etoolbox}
\newcommand{\MMtablefont}{\fontsize{9}{11}\selectfont}

\newcommand{\MMtablecaptionfont}{\fontsize{10}{12}\selectfont}
\newcommand{\MMtablerowstretch}{1.15}
\DeclareCaptionFont{mmtable}{\MMtablecaptionfont}
\AtBeginEnvironment{tabular}{\MMtablefont\renewcommand{\arraystretch}{\MMtablerowstretch}}
\AtBeginEnvironment{table}{\let\footnotesize\MMtablenotefont}
\newcommand{\MMtablefit}[2]{\adjustbox{max width=#1}{#2}}

%% file: sections/introduction_rev1.tex
\section{Introduction}\label{sec:introduction}

The Michaelis--Menten \rev{(MM)} model remains one of the standard tools of enzyme kinetics because it provides a parsimonious and interpretable relationship between substrate concentration and reaction velocity \cite{michaelis1913,johnson2011}. \rev{Recent applications include enzyme engineering with a self-driving laboratory \cite{rapp2024}, a protein-language-model-guided study of poly(ethylene terephthalate) (PET) hydrolases \cite{wu2025}, and soil exoenzyme studies across depths and temperatures \cite{alves2021}. The PET hydrolase study measured kinetics using the soluble substrate $p$-nitrophenyl butyrate, not insoluble PET. The Rapp and Alves studies also provide our data examples.} A large literature has therefore focused on how the two canonical kinetic parameters, the maximal reaction velocity $V_{\max}$ and the substrate concentration at half-maximal velocity $K_m$, should be estimated, from direct statistical fitting to graphical and distribution-free alternatives \cite{wilkinson1961,eisenthal1974,cornish1974,atkins1980}. \rev{Previous work emphasizes assessing uncertainty alongside estimates of $V_{\max}$ and $K_m$} \cite{mason2000}.

In practice, however, uncertainty quantification can be more fragile than point estimation. Confidence limits for $V_{\max}$ and especially $K_m$ depend on design quality and on how the error structure is handled \cite{cornish1972,duggleby1986,nimmo1974,matyska1985}. This matters in life-science studies where fitted kinetic parameters are compared across strains, temperatures, soils, or treatment groups, because underestimating uncertainty can distort downstream scientific interpretation even when the fitted curves look visually adequate. Routine analyses still often rely on homoscedastic nonlinear least squares even when assay variability increases with substrate concentration. In enzyme-kinetic applications, this kind of nonconstant variance has long motivated weighting and transformation strategies \cite{askelof1976,little1982,cornish1981,ruppert1989,alves2021}.

\begin{revision}
\label{par:existing-context}
Carroll and Ruppert \cite{carroll1988} describe mean- and covariate-dependent
variance functions, weighting, and transform-both-sides methods for nonlinear
regression (Supplementary Section~\ref*{supp-sec:existing}).
Transforming both the response and the fitted curve retains the kinetic
parameters, but estimating the mean response on the original scale may require
a correction. Davidian and Giltinan \cite{davidian1995} develop nonlinear
mixed-effects models for changing variance and dependence within subjects.
Existing R tools provide standard errors and confidence
intervals \cite{ritz2005,aledo2022};
\texttt{drc} supports weighting and transform-both-sides analysis
\cite{ritz2015}, and \texttt{nlme} supports generalized nonlinear least
squares (GNLS) and mixed-effects fitting \cite{nlme2026}. These tools can account
for changing variance. The remaining issue is therefore not whether
variance can be modeled, but whether estimation makes full use of the
information in the errors. When errors are asymmetric, variance weighting
alone may not fully use the information in the errors. Kim and Ma
\cite[Section~3, pp.~756--757]{kim2012} show that accounting for additional
features of the error distribution, such as asymmetry, can give more precise
estimates in large samples. This advantage requires correct assumptions
about the errors
\cite{kim2019}.\label{par:existing-context-end}
\end{revision}

\rev{To use this additional information,}
we adapt the heteroscedastic semiparametric score framework of Kim and Ma
\cite{kim2012} to Michaelis--Menten regression. \rev{We write
$\operatorname{Var}(\varepsilon\mid S)=\gamma h(S)$, with unknown scale
$\gamma>0$ and a fixed positive working variance function $h$.
\label{par:higher-moment-contribution}For independent curves, the estimating
equations use higher-order error moments to account for error asymmetry
beyond variance weighting. When the model and error moments are correct,
the estimator is semiparametrically efficient under standard regularity
conditions. Standard errors and confidence intervals are based on the
semiparametric theory of Kim and Ma \cite{kim2012} and Tsiatis
\cite{tsiatis2006}. With Gaussian
working moments, only $K_m$ requires numerical estimation, while
$V_{\max}$ and $\gamma$ have closed-form updates. User-defined $h$ is
also supported for single-curve and grouped analyses. For repeated-measurement
data, we extend the MM model to account for within-cluster dependence and
concentration-dependent residual variance \cite{davidian1995,magenton2010}.}

We study the resulting procedures in parallel single-curve and clustered simulation benchmarks, illustrate the method with two real enzyme data sets, and implement the workflow in a companion R package, \texttt{inferMM}, for single-curve fitting, variance-model screening, and grouped analysis. \rev{The package is available on CRAN. Its MM-specific interface reduces the setup needed for model fitting, variance-model comparison, and inferential summaries compared with general-purpose tools such as \texttt{nlme}.} Section~\ref{sec:methods} introduces the model and estimators, Section~\ref{sec:simulation} presents the simulation study, Section~\ref{sec:real-data} analyzes the real data, and Section~\ref{sec:discussion} concludes.

%% file: sections/methods_rev1.tex
\section{Semiparametric estimation for heteroscedastic and clustered Michaelis--Menten models}\label{sec:methods}

\subsection{Michaelis--Menten mean model}\label{sec:mm-mean-model}

We use the same Michaelis--Menten mean structure for both the single-curve and clustered settings. Following the classical formulation \cite{michaelis1913,johnson2011}, we write the conditional mean reaction velocity at substrate concentration $S$ as
\begin{equation}
\mu(S;V_{\max},K_m) = \frac{V_{\max} S}{K_m + S},
\label{eq:mm-mean}
\end{equation}
where $V_{\max}$ denotes the maximal reaction velocity, $K_m$ is the substrate concentration at which the mean velocity reaches one half of $V_{\max}$, and $\boldsymbol{\beta}=(V_{\max},K_m)^\top$. Equivalently,
\begin{equation}
Y_i = \mu(S_i;V_{\max},K_m) + \varepsilon_i,
\end{equation}
where $(S_i,Y_i)$ denotes the $i$th observation in the single-curve setting and cluster $i$ contains observations $(S_{ij},Y_{ij})$, $j=1,\ldots,n_i$, in the repeated-measurement setting. We retain this familiar mean structure and focus on extending the error specification to allow heteroscedasticity and clustering. \rev{We assume $V_{\max}>0$ and $K_m>0$.
To avoid confusion with substrate concentration $S$, we use $\psi$ for score functions in the estimating equations.}

\begin{figure}[htbp]
\centering
\includegraphics[width=\textwidth]{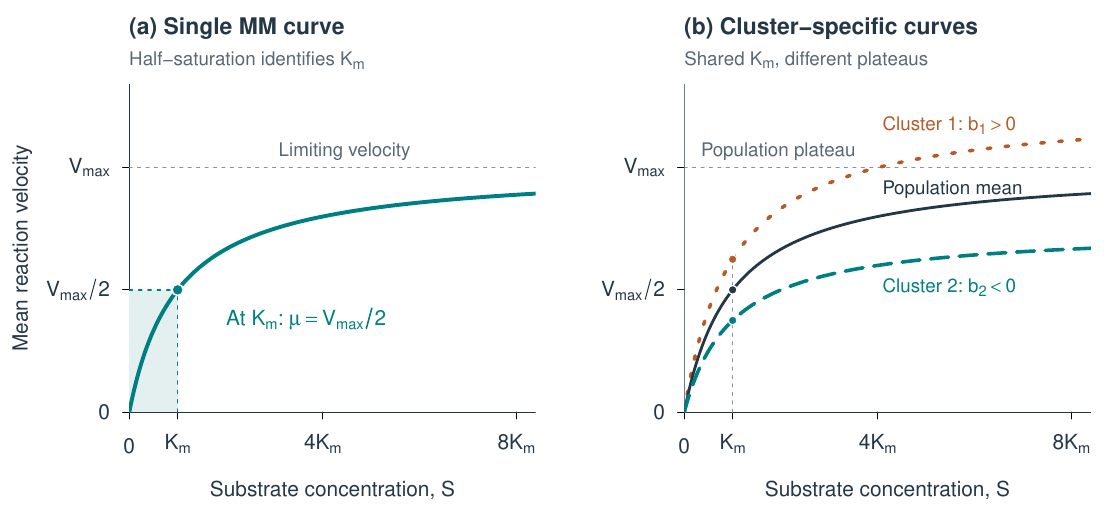}
\caption{\rev{Conceptual illustration of the Michaelis--Menten mean model.
(a) The mean reaches half its limiting reaction velocity $V_{\max}$ at
substrate concentration $K_m$.
(b) The dark solid curve is the population mean under
$E(b_i\mid\mathbf S_i)=0$. It approaches $V_{\max}$, marked by the horizontal
dotted line. All clusters share $K_m$, and cluster $i$ approaches
$V_{\max}+b_i$. Cluster 1 has $b_1>0$, so its curve can rise above the
population plateau. The schematic curves use arbitrary units and omit
residual noise.}}
\label{fig:mm-concept}
\end{figure}

\rev{Figure~\ref{fig:mm-concept} illustrates the meanings of these two
parameters and the cluster-specific kinetic scales introduced in
Section~\ref{sec:cluster-construction}.}

\subsection{Single-curve heteroscedastic model}\label{sec:single}

\subsubsection{Variance structure and semiparametric framework}

We first consider the single-curve case with independent observations $(S_i,Y_i)$, $i=1,\ldots,n$, and allow the random error to have mean zero but a variance that depends on substrate concentration. Specifically, we write
\begin{equation}
E(\varepsilon_i \mid S_i) = 0,\qquad
\operatorname{Var}(\varepsilon_i \mid S_i) = \gamma h(S_i),
\label{eq:variance-structure}
\end{equation}
where $h(\cdot)$ is a prespecified working variance function and $\gamma>0$ is an unknown scale parameter. \rev{We require $h(S_i)>0$; the constant-variance baseline is $h(S)\equiv1$. }
 This multiplicative form separates the overall variance level from the concentration-dependent shape, retains heteroscedasticity, and keeps the \rev{variance specification} low-dimensional \cite{kim2023}.

\rev{For fixed $h$, the mean and variance are specified by the three
parameters $(V_{\max},K_m,\gamma)$. The conditional error distribution is
left unspecified subject to those mean and variance restrictions, making
it the nonparametric component of the semiparametric model. The efficient
scores derived for this model involve conditional third and fourth error
moments \cite[Theorem~2]{kim2012} (Supplementary
Section~\ref*{supp-sec:supp-score}). To compute estimates, we can choose
a working error distribution, plug its moment formulas into the scores,
and solve the resulting estimating equations.}

\rev{For fixed $h$, substituting Gaussian error moments into the
Kim--Ma score gives the weighted nonlinear least-squares (WNLS) mean
equation described by Carroll and Ruppert~\cite{carroll1988}.
The default fit in \texttt{inferMM} solves this equation by profiling. The function \texttt{fit\_mm\_moments()} can use third and
fourth moments from non-Gaussian working distributions (Supplementary
Section~\ref*{supp-sec:supp-higher-moments}). With correct mean and
variance models, correctly specified or suitably estimated higher
moments, and standard regularity conditions, the resulting estimator
attains the semiparametric efficiency bound.}

\subsubsection{\rev{Gaussian-working profile-score estimation}}\label{sec:reduced-score}

\begin{revision}
An interior maximum likelihood estimate solves the log-likelihood score
equation. Likewise, we plug working moments into the semiparametric
efficient score \cite[Section~3.4]{tsiatis2006} and set the sample score
to zero. The Gaussian-working mean score for observation $i$
(Supplementary equation~(\ref*{supp-eq:reduced-scores})) is
\[
\psi_{\beta,\mathrm{eff}}(S_i,Y_i)
=\frac{Y_i-\mu(S_i;\boldsymbol\beta)}{\gamma h(S_i)}
 g(S_i;\boldsymbol\beta),
\qquad\text{where}\quad
g(S;\boldsymbol\beta)=\frac{\partial\mu(S;\boldsymbol\beta)}
{\partial\boldsymbol\beta}.
\]

At a trial $k=K_m$, the curve shape is fixed (Figure~\ref{fig:mm-concept}a)
and the mean is linear in its height $V=V_{\max}$, with
\[
\mu(S_i;V,k)=Vu_i(k),
\qquad\text{where}\quad u_i(k)=\frac{S_i}{k+S_i}.
\]
With $h_i=h(S_i)$ fixed, the weighted residual sum of squares and its
gradient are
\[
\begin{aligned}
Q(V,k)&=\sum_i\frac{r_i(V,k)^2}{h_i},
\qquad\text{where}\quad r_i(V,k)=Y_i-Vu_i(k),\\
\nabla_{\boldsymbol\beta}Q(V,k)
&=-2\gamma\sum_i\psi_{\beta,\mathrm{eff}}(S_i,Y_i).
\end{aligned}
\]
The gradient is zero when the sample mean score is zero. At fixed $k$,
the $V$-component gives the height equation and its solution,
\[
\begin{aligned}
\frac{\partial Q(V,k)}{\partial V}
&=-2\sum_i\frac{u_i(k)r_i(V,k)}{h_i}=0,\\
V(k)&=\frac{\sum_i u_i(k)Y_i/h_i}{b(k)},
\qquad\text{where}\quad b(k)=\sum_i\frac{u_i(k)^2}{h_i}.
\end{aligned}
\]
At each trial $k$, we have already fitted the height $V(k)$, so we only
need to see how the weighted residual sum of squares changes as we vary $k$.
The $V$-derivative vanishes at $V(k)$, so only the $k$-derivative contributes.
\begin{equation}
\begin{aligned}
F_h(k)&=-b(k)\sum_i\frac{S_i\,r_i(V(k),k)}{(k+S_i)^2h_i},\\
\frac{\mathrm{d}}{\mathrm{d}k}Q(V(k),k)
&=-\frac{2V(k)}{b(k)}F_h(k).
\end{aligned}
\label{eq:profile-Fh}
\end{equation}

The profile step substitutes the fitted height $V(k)$ into the
Gaussian-working $K_m$ score, leaving a function of $k$ alone.
For $V(k)>0$, the derivative in~\eqref{eq:profile-Fh} is zero exactly
when $F_h(k)=0$. We search for a positive root $\hat K_m$ that passes
the checks in Section~\ref{sec:profile} and obtain the fitted height as
\begin{equation}
\hat V_{\max}=V(\hat K_m).
\label{eq:vmax-plugin}
\end{equation}
The fitted residuals then give the scale estimate
\begin{equation}
\hat\gamma=
\frac{1}{n}\sum_{i=1}^n\frac{\hat\varepsilon_i^2}{h_i},
\qquad\text{where}\quad
\hat\varepsilon_i=Y_i-\frac{\hat V_{\max}S_i}{\hat K_m+S_i}.
\label{eq:gamma-plugin}
\end{equation}
The fitted variance is $\hat\gamma h_i$. Supplementary
Section~\ref*{supp-sec:supp-profile-details} gives the full derivation.
\end{revision}

\subsubsection{Inference and implementation}\label{sec:profile}

\begin{revision}
To quantify how parameter estimates vary across assays at the same
concentrations, we use the mean gradient $g(S;\boldsymbol\beta)$ from
Section~\ref{sec:reduced-score}. The mean response is most sensitive to
$V_{\max}$ near the plateau and to $K_m$ near $S=K_m$. For the same
gradient, noisier observations contribute less information
(Supplementary Section~\ref*{supp-sec:supp-gaussian-moments}).

We plug $\hat{\boldsymbol\beta}=(\hat V_{\max},\hat K_m)^\top$ into the
gradients and average their outer products with weights $1/h_i$,
\[
\widehat M_{\beta\beta}^{(h)}
=
\frac{1}{n}\sum_{i=1}^{n}
\frac{g(S_i;\hat{\boldsymbol{\beta}})g(S_i;\hat{\boldsymbol{\beta}})^\top}{h(S_i)}.
\]
Using the scale $\hat\gamma$ from~\eqref{eq:gamma-plugin}, we invert
the total working information $n\widehat M_{\beta\beta}^{(h)}/\hat\gamma$
to obtain the plug-in covariance estimate~\cite[Chapter~4]{tsiatis2006},
\begin{equation}
\widehat{\operatorname{Var}}(\hat{\boldsymbol{\beta}})
=
\frac{\hat\gamma}{n}
\left(\widehat M_{\beta\beta}^{(h)}\right)^{-1}.
\label{eq:profile-covariance}
\end{equation}

The square roots of the diagonal entries of~\eqref{eq:profile-covariance}
give standard errors for the default Wald intervals, based on asymptotic
normality under the stated model and regularity conditions~\cite{kim2012}.
Supplementary Section~\ref*{supp-sec:supp-covariance} describes sandwich
and HC3 comparisons.
\end{revision}

\rev{For higher-moment estimation of independent curves,%
\label{par:covariance-comparison}\label{sec:higher-moments}
the default covariance uses the inverse joint working information for
$(V_{\max},K_m,\gamma)$. Efficiency requires correct mean, variance, and
error moments and standard regularity conditions. For the same fit,
the optional sandwich uses derivatives and observed variability of the
estimating equations to calculate standard errors and intervals
(Supplementary Sections~\ref*{supp-sec:supp-general-moments}
and~\ref*{supp-sec:supp-higher-moments}). Clustered inference is described
in Section~\ref{sec:cluster-algorithm}.}

\Needspace{9\baselineskip}
\begin{revision}
We check score convergence, parameter validity, and numerical stability
before reporting results (Supplementary
Section~\ref*{supp-sec:supp-profile-details}).
\phantomsection\label{sec:bootstrap}The optional studentized wild bootstrap
for independent single curves uses Mammen multipliers on fitted residuals
and refits at fixed concentrations to calibrate intervals (Supplementary
Section~\ref*{supp-sec:supp-bootstrap}). Working-Gaussian AIC and BIC
compare fixed variance shapes, and intervals condition on the selected shape
(Supplementary Section~\ref*{supp-sec:supp-screening}). Supplementary
Section~\ref*{supp-sec:supp-higher-benchmark} compares higher-moment
estimation with WNLS.
\end{revision}

\Needspace{9\baselineskip}
\subsection{Clustered extension}\label{sec:cluster}

\subsubsection{Working covariance and score construction}\label{sec:cluster-construction}

\begin{revision}
Readings from the same assay run, plate or biological sample form a cluster
when they share random variation; different clusters are independent.
Conditionally independent replicates can instead use the single-curve model.

For cluster $i=1,\ldots,m$, let $j=1,\ldots,n_i$ index its $n_i$
readings and $\mathbf S_i=(S_{i1},\ldots,S_{in_i})^\top$ contain their substrate
concentrations. A shared random effect on $V_{\max}$ gives
\cite{davidian1995}
\begin{equation}
Y_{ij}=(V_{\max}+b_i)\frac{S_{ij}}{K_m+S_{ij}}+\varepsilon_{ij},
\qquad E(b_i\mid\mathbf S_i)=0,\quad
\operatorname{Var}(b_i\mid\mathbf S_i)=\tau^2.
\label{eq:cluster-model}
\end{equation}
where $b_i$ describes between-cluster activity differences.
\end{revision}
\begin{revision}
Define
\[
z_{ij}=\frac{S_{ij}}{K_m+S_{ij}},
\qquad
\mathbf z_i=(z_{i1},\ldots,z_{in_i})^\top,
\qquad
A_i=\operatorname{diag}\{h(S_{ij})\}_{j=1}^{n_i}.
\]
Conditional residuals have mean zero and covariance $\gamma A_i$ given
$b_i$ and $\mathbf S_i$. Thus the response vector
$\mathbf Y_i=(Y_{i1},\ldots,Y_{in_i})^\top$ has mean
$\boldsymbol\mu_i(\boldsymbol\beta)=V_{\max}\mathbf z_i$ and covariance
\begin{equation}
V_i=\tau^2 \mathbf{z}_i \mathbf{z}_i^\top+\gamma A_i,
\label{eq:cluster-covariance}
\end{equation}
The first term represents shared activity variation and the second residual
noise. The loading matrix $Z_i=\mathbf z_i(K_m)$ changes with $K_m$;
its contribution $z_{ij}b_i$ is small at low concentrations and approaches
$b_i$ at saturation.
\end{revision}

\begin{revision}
Let $D_i(\boldsymbol\beta)=\partial\boldsymbol\mu_i(\boldsymbol\beta)/
\partial\boldsymbol\beta^\top$ be the mean-gradient matrix and
$\mathbf r_i=\mathbf Y_i-\boldsymbol\mu_i(\boldsymbol\beta)$ the residual
vector. A generalized estimating equation (GEE)-style mean-only working score is
\cite[Section~3.1, equation~(6)]{liang1986}
\end{revision}
\begin{equation}
\rev{\psi_{i,\beta}^{\mathrm{rep}}}
=
D_i(\boldsymbol{\beta})^\top
V_i^{-1}
\{\mathbf{Y}_i-\boldsymbol{\mu}_i(\boldsymbol{\beta})\}.
\end{equation}
\begin{revision}
When $V_i$ depends on $\beta_a$, the full Gaussian score also includes
its covariance derivative:
\begin{equation}
D_{i,a}^\top V_i^{-1}\mathbf r_i+
\frac12\{\mathbf r_i^\top V_i^{-1}V_{i,a}V_i^{-1}\mathbf r_i
-\operatorname{tr}(V_i^{-1}V_{i,a})\},
\label{eq:cluster-gaussian-score}
\end{equation}
where $D_{i,a}$ is column $a$ of $D_i$ and
$V_{i,a}=\partial V_i/\partial\beta_a$. Because $V_i$ depends on $K_m$,
the default profile below retains this covariance dependence rather than
solving the mean-only equation alone.
\end{revision}

\begin{revision}
\paragraph{Ma--Genton-style latent-variable formulation}
Ma and Genton \cite[Sections~1 and 3.2]{magenton2010} leave the latent
distribution unspecified while modelling conditional responses, including
gamma and Poisson families \cite[Section~5]{magenton2010}.
Our clustered \texttt{projected} option implements the conditionally
Gaussian MM case. With $T_i=V_{\max}+b_i$, model~\eqref{eq:cluster-model}
becomes
\[
\begin{aligned}
Y_{ij}\mid T_i,\mathbf S_i
&\overset{\mathrm{ind}}{\sim}
N\!\left(T_i\frac{S_{ij}}{K_m+S_{ij}},\,\gamma h(S_{ij})\right),\\
E(T_i\mid\mathbf S_i)&=V_{\max},\qquad
\operatorname{Var}(T_i\mid\mathbf S_i)=\tau^2.
\end{aligned}
\]
Measurement errors within a sample may therefore be Gaussian while
activity differences between samples are skewed: the distribution of $T_i$
is unspecified. We call this option \texttt{projected} because a weighted
projection separates the curve-height component along $\mathbf z_i(k)$
from the orthogonal residual used in the $K_m$ estimating equation.
Other conditional response families require separate
equations and are not yet implemented in this option. For the profile
method, ``Gaussian-working'' describes its fitting criterion, not a
requirement of exactly Gaussian activity. Table~\ref{tab:cluster-paths}
compares the methods; derivations and efficiency scope are given in
Supplementary Section~\ref*{supp-sec:supp-cluster-projected}.
\end{revision}

\begin{table}[!htbp]
\centering
\ifshowrevisions\color{blue}\fi
\caption{\rev{Clustered estimation paths in \texttt{inferMM}. Both use
the same MM mean and concentration-dependent residual variance shape.}}
\label{tab:cluster-paths}
\small
\setlength{\tabcolsep}{4pt}
\renewcommand{\arraystretch}{1.15}
\begin{tabular}{@{}>{\raggedright\arraybackslash}p{0.22\linewidth}>{\raggedright\arraybackslash}p{0.36\linewidth}>{\raggedright\arraybackslash}p{0.36\linewidth}@{}}
\toprule
Feature & Gaussian-working profile/moment & Ma--Genton-style projected score \\
\midrule
Interface & \texttt{method="profile"}\newline (default)
& \texttt{method="projected"}\newline (optional) \\[4pt]
Random-effect treatment & Gaussian-working criterion; variance components estimated by moments
& Density unspecified; conditionally Gaussian measurement errors \\[4pt]
Mean estimation & Fits with cluster covariance; alternates mean and variance updates
& Separates activity-level differences; solves a different $K_m$ equation \\[4pt]
Suitable use & Default covariance-based clustered analysis
& Unknown activity distribution with plausibly Gaussian within-sample errors \\[4pt]
Concentration layout & Replicated clusters; concentrations may vary across or within clusters
& Within-cluster variation at positive concentrations is required \\[4pt]
Uncertainty & Mean-information working approximation
& Stacked independent-cluster sandwich \\
\bottomrule
\end{tabular}
\end{table}
\FloatBarrier

\subsubsection{Estimator and inference}\label{sec:cluster-algorithm}

\rev{The default \texttt{method="profile"} starts from the pooled
single-curve fit, then alternates a scalar $K_m$ update, a plug-in
$V_{\max}$ update and moment updates for $(\tau^2,\gamma)$.
The reported numerical analyses use this profile/moment procedure.}

\begin{revision}
With $\tau^2$ and $\gamma$ fixed during a mean update, define
\begin{align}
V(k)&=\frac{\sum_i\mathbf z_i(k)^\top V_i(k)^{-1}\mathbf Y_i}
{\sum_i\mathbf z_i(k)^\top V_i(k)^{-1}\mathbf z_i(k)},\nonumber\\
L_p(k)&=\sum_i[\log|V_i(k)|+
\mathbf r_i(k)^\top V_i(k)^{-1}\mathbf r_i(k)],
\label{eq:cluster-profile}
\end{align}
where $\mathbf r_i(k)=\mathbf Y_i-V(k)\mathbf z_i(k)$.
Minimize $L_p(k)$ on a positive log-scale interval and substitute into
$V(k)$. For current residuals $r_{ij}$, the unconstrained moment updates are
\begin{equation}
\tilde\tau^2=
\frac{\sum_i\sum_{j<l}z_{ij}z_{il}r_{ij}r_{il}}
{\sum_i\sum_{j<l}(z_{ij}z_{il})^2},\qquad
\tilde\gamma=
\frac{\sum_{ij}h(S_{ij})\{r_{ij}^2-\hat\tau^2z_{ij}^2\}}
{\sum_{ij}h(S_{ij})^2}.
\label{eq:cluster-moments}
\end{equation}
Variance components are floored at $10^{-8}$. Alternation stops when the
largest mean-parameter change is below $10^{-5}$, with at most eight
iterations; reaching the limit retains the fit with a convergence warning.
The pooled initializer must pass the single-curve checks in Supplementary
Section~\ref*{supp-sec:supp-profile-details}.
\end{revision}

\rev{The mean-information covariance approximation is}
\begin{equation}
\widehat{\operatorname{Var}}(\hat{\boldsymbol{\beta}}_{\mathrm{rep}})
=
\left\{
\sum_{i=1}^m
D_i(\hat{\boldsymbol{\beta}}_{\mathrm{rep}})^\top
\widehat V_i^{-1}
D_i(\hat{\boldsymbol{\beta}}_{\mathrm{rep}})
\right\}^{-1}.
\label{eq:cluster-working-covariance}
\end{equation}
\rev{It conditions on the fitted covariance, omitting covariance derivatives
and joint variance-component uncertainty; inversion uses diagonal
regularization starting at $10^{-8}$. This hybrid profile/moment estimator
is not joint maximum likelihood or efficient-score estimation.
Its Wald calibration is assessed in the simulations.}

\begin{revision}
\paragraph{Ma--Genton-style projected-score option}
With \texttt{method="projected"}, conditional centering
\cite[Section~3.2]{magenton2010} separates sample-level activity from the
within-sample residual pattern. A scalar equation estimates the common
$K_m$, followed by explicit moment updates for $V_{\max}$, $\gamma$ and
$\tau^2$. A joint cluster sandwich combines the mean and variance equations
to give standard errors and normal-cutoff Wald intervals, rather than the
working approximation in \eqref{eq:cluster-working-covariance}.
Full equations and numerical checks are in Supplementary
Section~\ref*{supp-sec:supp-cluster-projected}.
\end{revision}

\begin{revision}
\paragraph{Method choice and assay examples}
Both methods can fit independent soil samples measured across a
concentration series. If most samples have moderate activity but a few
consistently show higher responses, \texttt{projected} offers estimation
and joint uncertainty calculations without choosing a parametric activity
distribution, under the stated conditional-Gaussian model and regularity
conditions. If instead independent samples are randomly assigned to one
concentration each, with technical repeats and several samples per
concentration, only \texttt{profile} applies: concentration differences
occur between samples, whereas projection requires variation within them.
The profile method also permits singleton samples alongside replicated
samples.
\end{revision}

\begin{revision}
Both routines require at least three distinct concentrations overall. Projection
additionally requires at least five clusters, two readings per cluster,
a positive concentration in every cluster and informative within-cluster
concentration variation; its joint covariance requires finite fourth
random-effect moments. These are eligibility conditions, not guarantees
of interval accuracy with few clusters.
\end{revision}

%% file: sections/simulation_design_rev1.tex
\section{Simulation study}\label{sec:simulation}
\subsection{Single-curve simulation benchmark}\label{sec:single-sim}
\subsubsection{Design}\label{sec:single-sim-design}

We first evaluated the independent single-curve estimator on a fixed-grid benchmark. The conditional mean followed \eqref{eq:mm-mean} with true parameter values $V_{\max}=100$ and $K_m=20$. The main-text benchmark used 50 fixed substrate concentrations equally spaced over $[1,100]$, yielding $n=50$ observations in each simulated data set, and each setting was replicated 1,000 times. \rev{In a relatively high-throughput microfluidic assay, Neun et al.\ \cite[Figure~3]{neun2022} fitted a Michaelis--Menten curve for the glycosidase SN243 using initial rates at 60 distinct substrate concentrations in one droplet series. Marcel et al.\ \cite[pp.~330--331]{marcel1998} used 48 distinct substrate concentrations in their cholinesterase assays. These examples motivate the number of concentrations in our benchmark, which compares variance shapes on a fixed grid with independent errors. Section~\ref{sec:additional-sim} covers smaller designs with 20 equally spaced concentrations and an SDL-inspired seven-concentration subset with four independently generated responses per level.}

For each simulation scenario, data were generated from the heteroscedastic Michaelis--Menten model
\begin{equation}
Y_i = \mu(S_i;\boldsymbol{\beta}_0) + \varepsilon_i, \qquad
\varepsilon_i = \sqrt{v(S_i)}\, z_i,
\end{equation}
where \label{par:single-error-distribution}$\boldsymbol{\beta}_0=(100,20)^\top$ and $z_i$ satisfies $E(z_i)=0$ and $\operatorname{Var}(z_i)=1$, so that the conditional variance of $Y_i$ given $S_i$ is exactly $v(S_i)$. \rev{The main-text fixed-grid results use independent $z_i\sim N(0,1)$. The parallel skewed-error benchmark uses $z_i=(G_i-4)/2$, with independent $G_i\sim\operatorname{Gamma}(4,1)$ of unit scale.}

To reflect plausible heteroscedastic patterns in enzyme-kinetic applications, we considered three increasing but bounded variance functions:
\begin{align}
v_{\mathrm{MM}}(s) &= 1 + 9 \frac{s}{20+s}, \nonumber\\
v_{\mathrm{Exp}}(s) &= 1 + 9 \left(1-e^{-0.05s}\right), \nonumber\\
v_{\mathrm{Hill}}(s) &= 1 + 9 \frac{s^2}{20^2+s^2}.
\label{eq:generating-variances}
\end{align}
The first specification represents a \rev{Michaelis--Menten-type saturation pattern (MM)} in the variance, the second gives a smoother \rev{exponential approach (Exp)} to an upper bound, and the third allows a more flexible \rev{Hill-type sigmoidal increase (Hill)}. \rev{Supplementary Figure~\ref*{supp-fig:variance-functions} plots all three generating functions, which differ from the fitted working variance shapes $h$.} All three variance functions are increasing and bounded, with a common baseline variance of 1 and an upper limit of 10, so that the comparison focuses on differences in variance shape rather than on large differences in overall noise magnitude.

To assess robustness against departure from the Gaussian working model, the Supplementary Material also reports a parallel skewed-error benchmark based on a standardized right-skewed innovation law. The \rev{additional generating details}, together with the corresponding single-curve skewed-error results, are deferred to the Supplementary Material, in particular Supplementary Table~\rev{\ref*{supp-tab:supp-main-point-skewed}} and Supplementary Table~\rev{\ref*{supp-tab:supp-main-diagnostic-skewed}}.

\subsubsection{Competing estimators and performance measures}

In the \rev{$n=50$ fixed-grid} benchmark, we compared standard nonlinear least squares (NLS) and three \rev{Gaussian-working profile} estimators with fixed working variance shapes $h(s)=\log(s+1)$, $h(s)=s^{1/2}$, and $h(s)=s^{1/3}$. NLS serves as the conventional homoscedastic baseline. The three \rev{profile} competitors differ only in the choice of $h(\cdot)$, so the comparison isolates whether simple low-dimensional variance models can improve estimation and inference for $V_{\max}$ and $K_m$ when the true heteroscedastic structure is unknown. \rev{The table label \emph{Profile} identifies the Gaussian-working reduction used in these comparisons.}

These working variance functions were chosen as simple monotone approximations to plausible concentration-dependent noise patterns. For a fixed $h(\cdot)$, the resulting mean estimator already matches the corresponding weighted nonlinear least-squares score equations, so a separate fixed-weight WNLS row would mainly duplicate the same benchmark under a different label. \rev{For the Grid20 and SDL28 designs, Section~\ref{sec:additional-sim} adds fitted-power GNLS, HC3 standard errors, and AIC screening over the three shapes above and $h(s)\equiv1$. The exploratory SDL28 and SDL56 comparison adds a shape proportional to the squared pilot mean as a fifth candidate (Supplementary Tables~\mbox{\ref*{supp-tab:matched-transform-point}--\ref*{supp-tab:matched-transform-inference}}). Supplementary Tables \ref*{supp-tab:matched-cluster-point}--\ref*{supp-tab:matched-paired} additionally examine a matched-covariance NLME comparison. The direct higher-moment comparison and the complete Gaussian-working/log-TBS sensitivity analysis are given in Supplementary Section~\ref*{supp-sec:supp-matched}.}

\rev{The fixed-grid analyses use the estimation routines distributed in \texttt{inferMM} 0.1.2, with seed 2026. The NLS comparator uses the covariance returned by \texttt{nls()}, including residual degrees-of-freedom scaling, whereas the profile methods use the $n^{-1}$ scale update. All use normal cutoffs in this simulation. These covariance conventions can affect finite-sample coverage independently of the weights.}

%% file: sections/additional_simulation_rev1.tex
\subsection{\rev{Small-sample, replicated-design, and constant-variance checks}}
\label{sec:additional-sim}
\begin{revision}
To assess smaller assays, we simulated 20 equally spaced concentrations
and an SDL-inspired seven-concentration subset design with four independently
generated responses per level (28 observations). The full SDL application
uses eight concentrations; these simulations assess the stated independent-error
design rather than verify independence in the observed assays. We considered the
three variance patterns in \eqref{eq:generating-variances} and a
constant-variance baseline, under normal and moderately skewed errors,
with 1,000 datasets per setting. Comparisons included
fixed-shape profiles, AIC variance selection, HC3 standard errors, and
GNLS with an estimated variance exponent.

At $n=20$, AIC screening yielded lower RMSEs for both parameters than
the unweighted fit in all six heteroscedastic settings, and the unweighted
fit had the lowest RMSEs under constant variance. Empirical coverage of
the screen's nominal 95\% Wald intervals was 89.2--93.7\%
(Supplementary Tables~\mbox{\ref*{supp-tab:revision-Grid20-Normal}--\ref*{supp-tab:revision-Grid20-Skewed}}).
Replicate variability and residual checks help guide variance-model choice.
Supplementary Section~\ref*{supp-sec:supp-benchmark}
(Tables~\mbox{\ref*{supp-tab:revision-homoscedastic}--\ref*{supp-tab:revision-SDL28-Skewed}})
gives the designs and full results.
Supplementary Section~\ref*{supp-sec:supp-matched} gives matched WNLS and
NLME comparisons and the comparison of Gaussian-working profiling with
log-TBS fitted in \texttt{drc} (Section~\ref*{supp-sec:supp-tbs-sensitivity},
Tables~\mbox{\ref*{supp-tab:matched-transform-point}--\ref*{supp-tab:matched-transform-inference}}).
In the exploratory SDL28 and SDL56 comparison, the expanded AIC screen
added $h(s)=\{s/(\widehat K_{m,0}+s)\}^2$, where $\widehat K_{m,0}$ is from
the pilot profile fit with $h(s)=s^{1/2}$. Under lognormal errors, the expanded
screen reduced both kinetic-parameter RMSEs and the variance RMSE relative
to the original screen.
With HC3 standard errors and $t_{n-2}$ cutoffs, SDL28 $V_{\max}$ coverage
was 93.5\% for additive errors and 94.7\% for lognormal errors, both below
the nominal 95\%.
\end{revision}

%% file: sections/discussion_rev1.tex
\section{Discussion and conclusion}\label{sec:discussion}

\rev{When assay variability changes with concentration, kinetic comparisons benefit from accounting for unequal measurement precision. Our framework provides this adjustment on the original response scale through conditional-moment modeling. Carroll and Ruppert~\cite{carroll1988} also cover variance functions and transform-both-sides models. Both approaches retain kinetic parameters, but transformation models specify errors on a transformed scale and may require retransformation correction for the original-scale conditional mean. In our fixed log-TBS comparison, square-root profiling had lower parameter RMSEs under additive errors and log-TBS under lognormal errors (Supplementary Tables~\ref*{supp-tab:matched-transform-point}--\ref*{supp-tab:matched-transform-inference}). This supports assay-specific variance assessment using residual diagnostics and working AIC/BIC, including a constant-variance candidate.}

\rev{Variance weighting uses differences in measurement precision when fitting the curve. With fixed weights, the Gaussian-working profile has the same mean equations as WNLS, and its MM-specific reduction leaves only $K_m$ for numerical search. Under regularity conditions, sandwich covariance estimates the asymptotic covariance of a given estimator, providing standard errors robust to misspecification of the working variance function~\cite{zeileis2006}. The two can be combined, as in our HC3 analysis. Screened intervals condition on the selected variance shape and, for the expanded screen, the pilot curve.}

\rev{The higher-moment extension uses conditional third and fourth moments to exploit asymmetry, following Kim and Ma~\cite{kim2012}. The 25--31\% sampling-variance reductions relative to WNLS in the Gamma comparison illustrate gains from this additional information (Supplementary Tables~\ref*{supp-tab:higher-moment-point}--\ref*{supp-tab:higher-moment-inference}). Semiparametric efficiency requires correct mean and variance specifications, suitable moment estimation and regularity. Reliable higher-moment estimation may require larger samples.}

\rev{The clustered extension separates between-cluster variation from concentration-dependent residual noise, with point accuracy comparable to NLME~\cite{davidian1995}. With six clusters, default-profile $V_{\max}$ coverage was 85.5\%, and NLME showed similar undercoverage. The profile covariance conditions on fitted variance components, motivating further assessment of joint uncertainty and small-cluster interval calibration. In the pooled Alves analysis, concentration dependence and between-condition differences both contribute to variability. A hierarchical analysis retaining depth, temperature and core effects would support condition-specific interpretation.}

%% file: sections/figure_legends_rev1.tex
\clearpage
\section*{Figure legends}

\noindent\textbf{Figure 1.} \rev{Conceptual illustration of the Michaelis--Menten mean model.
(a) The mean reaches half its limiting reaction velocity $V_{\max}$ at
substrate concentration $K_m$.
(b) The dark solid curve is the population mean under
$E(b_i\mid\mathbf S_i)=0$. It approaches $V_{\max}$, marked by the horizontal
dotted line. All clusters share $K_m$, and cluster $i$ approaches
$V_{\max}+b_i$. Cluster 1 has $b_1>0$, so its curve can rise above the
population plateau. The schematic curves use arbitrary units and omit
residual noise.}
\par\medskip

\noindent\textbf{Figure 2.} Fitted Michaelis--Menten curves for the best-fitting model [$S^{1/2}$] across the six SDL enzyme variants (1111--6666). The black line shows the fitted mean curve; the shaded region represents the \rev{Gaussian-working residual envelope conditional on the fitted parameters, clipped below at zero}. Points indicate individual replicate observations\rev{, with 32 observations at eight concentrations per variant}. \rev{Velocities are in $\mu$M/min.}
\par\medskip

\noindent\textbf{Figure 3.} Fitted Michaelis--Menten curves for the best-fitting model [$S^{1/2}$] for AP, BG, and LAP in the Alves et al. (2021) data set. The shaded band expands with increasing substrate concentration, proportionally reflecting the heteroscedastic variance structure. \rev{The residual envelope is conditional on the fitted parameters and clipped at zero.} Points indicate pooled replicate observations across depth layers and temperatures.
\par\medskip

%% file: references.bib
@article{michaelis1913,
  author = {Michaelis, L. and Menten, M. L.},
  title = {Die Kinetik der Invertinwirkung},
  journal = {Biochem. Z.},
  volume = {49},
  pages = {333--369},
  year = {1913}
}

@article{wilkinson1961,
  author = {Wilkinson, G. N.},
  title = {Statistical estimations in enzyme kinetics},
  journal = {Biochem. J.},
  volume = {80},
  number = {2},
  pages = {324--332},
  year = {1961},
  doi = {10.1042/BJ0800324}
}

@article{johnson2011,
  author = {Johnson, Kenneth A. and Goody, Roger S.},
  title = {The original {Michaelis} constant: translation of the 1913 {Michaelis--Menten} paper},
  journal = {Biochemistry},
  volume = {50},
  number = {39},
  pages = {8264--8269},
  year = {2011},
  doi = {10.1021/bi201284u}
}

@article{cornish1972,
  author = {Cornish-Bowden, A. J.},
  title = {Analysis of progress curves in enzyme kinetics},
  journal = {Biochem. J.},
  volume = {130},
  number = {2},
  pages = {637--639},
  year = {1972},
  doi = {10.1042/BJ1300637}
}

@article{eisenthal1974,
  author = {Eisenthal, R. and Cornish-Bowden, A.},
  title = {The direct linear plot: {A} new graphical procedure for estimating enzyme kinetic parameters},
  journal = {Biochem. J.},
  volume = {139},
  number = {3},
  pages = {715--720},
  year = {1974},
  doi = {10.1042/BJ1390715}
}

@article{cornish1974,
  author = {Cornish-Bowden, A. and Eisenthal, R.},
  title = {Statistical considerations in the estimation of enzyme kinetic parameters by the direct linear plot and other methods},
  journal = {Biochem. J.},
  volume = {139},
  number = {3},
  pages = {721--730},
  year = {1974},
  doi = {10.1042/BJ1390721}
}

@article{nimmo1974,
  author = {Nimmo, I. A. and Atkins, G. L.},
  title = {A comparison of two methods for fitting the integrated {Michaelis--Menten} equation},
  journal = {Biochem. J.},
  volume = {141},
  number = {3},
  pages = {913--914},
  year = {1974},
  doi = {10.1042/BJ1410913}
}

@article{atkins1980,
  author = {Atkins, G. L. and Nimmo, I. A.},
  title = {Current trends in the estimation of {Michaelis--Menten} parameters},
  journal = {Anal. Biochem.},
  volume = {104},
  number = {1},
  pages = {1--9},
  year = {1980},
  doi = {10.1016/0003-2697(80)90268-7}
}

@article{cornish1981,
  author = {Cornish-Bowden, A. and Endrenyi, L.},
  title = {Fitting of enzyme kinetic data without prior knowledge of weights},
  journal = {Biochem. J.},
  volume = {193},
  number = {3},
  pages = {1005--1008},
  year = {1981},
  doi = {10.1042/BJ1931005}
}

@article{askelof1976,
  author = {Askel{\"o}f, Per and Korsfeldt, Margareta and Mannervik, Bengt},
  title = {Error structure of enzyme kinetic experiments: implications for weighting in regression analysis of experimental data},
  journal = {Eur. J. Biochem.},
  volume = {69},
  number = {1},
  pages = {61--67},
  year = {1976},
  doi = {10.1111/j.1432-1033.1976.tb10858.x}
}

@article{little1982,
  author = {Little, D. I. and Poat, P. C. and Giles, I. G.},
  title = {Residual analysis in determining the error structure in enzyme kinetic data: simulation experiments and observations on {Carcinus maenas} phosphofructokinase},
  journal = {Eur. J. Biochem.},
  volume = {124},
  number = {3},
  pages = {499--505},
  year = {1982},
  doi = {10.1111/j.1432-1033.1982.tb06621.x}
}

@article{matyska1985,
  author = {Matyska, L. and Kov{\'a}{\v r}, J.},
  title = {Comparison of several non-linear-regression methods for fitting the {Michaelis--Menten} equation},
  journal = {Biochem. J.},
  volume = {231},
  number = {1},
  pages = {171--177},
  year = {1985},
  doi = {10.1042/BJ2310171}
}

@article{duggleby1986,
  author = {Duggleby, R. G.},
  title = {Progress-curve analysis in enzyme kinetics: {N}umerical solution of integrated rate equations},
  journal = {Biochem. J.},
  volume = {235},
  number = {2},
  pages = {613--615},
  year = {1986},
  doi = {10.1042/BJ2350613}
}

@article{mason2000,
  author = {Mason, Graeme F. and Lai, James C. K.},
  title = {Nonlinear determination of {Michaelis--Menten} kinetics with model evaluation through estimation of uncertainties},
  journal = {Metab. Brain Dis.},
  volume = {15},
  number = {2},
  pages = {133--149},
  year = {2000},
  doi = {10.1007/BF02679980}
}

@article{alves2021,
  author = {Alves, Ricardo J. Eloy and Callejas, Ileana A. and Marschmann, Gianna L. and Mooshammer, Maria and Singh, Hans W. and Whitney, Bizuayehu and Torn, Margaret S. and Brodie, Eoin L.},
  title = {Kinetic properties of microbial exoenzymes vary with soil depth but have similar temperature sensitivities through the soil profile},
  journal = {Front. Microbiol.},
  volume = {12},
  pages = {735282},
  year = {2021},
  doi = {10.3389/fmicb.2021.735282}
}

@article{ruppert1989,
  author = {Ruppert, David and Cressie, Noel and Carroll, Raymond J.},
  title = {A transformation/weighting model for estimating {Michaelis--Menten} parameters},
  journal = {Biometrics},
  volume = {45},
  number = {2},
  pages = {637--656},
  year = {1989},
  doi = {10.2307/2531506}
}

@article{kim2012,
  author = {Kim, Mijeong and Ma, Yanyuan},
  title = {The efficiency of the second-order nonlinear least squares estimator and its extension},
  journal = {Ann. Inst. Stat. Math.},
  volume = {64},
  number = {4},
  pages = {751--764},
  year = {2012},
  doi = {10.1007/s10463-011-0332-y}
}

@article{kim2019,
  author = {Kim, Mijeong and Ma, Yanyuan},
  title = {Semiparametric efficient estimators in heteroscedastic error models},
  journal = {Ann. Inst. Stat. Math.},
  volume = {71},
  number = {1},
  pages = {1--28},
  year = {2019},
  doi = {10.1007/s10463-017-0622-0}
}

@article{kim2023,
  author = {Kim, Mijeong},
  title = {Appropriate use of parametric and nonparametric methods in estimating regression models with various shapes of errors},
  journal = {Stat},
  volume = {12},
  number = {1},
  pages = {e606},
  year = {2023},
  doi = {10.1002/sta4.606}
}

@book{tsiatis2006,
  author = {Tsiatis, Anastasios A.},
  title = {Semiparametric Theory and Missing Data},
  series = {Springer Series in Statistics},
  publisher = {Springer},
  address = {New York},
  year = {2006},
  doi = {10.1007/0-387-37345-4}
}

@article{liang1986,
  author = {Liang, Kung-Yee and Zeger, Scott L.},
  title = {Longitudinal data analysis using generalized linear models},
  journal = {Biometrika},
  volume = {73},
  number = {1},
  pages = {13--22},
  year = {1986},
  doi = {10.1093/biomet/73.1.13}
}

@article{magenton2010,
  author = {Ma, Yanyuan and Genton, Marc G.},
  title = {Explicit estimating equations for semiparametric generalized linear latent variable models},
  journal = {Journal of the Royal Statistical Society: Series B (Statistical Methodology)},
  volume = {72},
  number = {4},
  pages = {475--495},
  year = {2010},
  doi = {10.1111/j.1467-9868.2010.00741.x}
}

@article{marcel1998,
  author = {Marcel, V{\'e}ronique and Palacios, Laurent Gagnoux and Pertuy, Christophe and Masson, Patrick and Fournier, Didier},
  title = {Two invertebrate acetylcholinesterases show activation followed by inhibition with substrate concentration},
  journal = {Biochem. J.},
  volume = {329},
  number = {2},
  pages = {329--334},
  year = {1998},
  doi = {10.1042/bj3290329}
}

@article{neun2022,
  author = {Neun, Stefanie and van Vliet, Liisa and Hollfelder, Florian and Gielen, Fabrice},
  title = {High-Throughput Steady-State Enzyme Kinetics Measured in a Parallel Droplet Generation and Absorbance Detection Platform},
  journal = {Anal. Chem.},
  volume = {94},
  number = {48},
  pages = {16701--16710},
  year = {2022},
  doi = {10.1021/acs.analchem.2c03164}
}

@article{rapp2024,
  author = {Rapp, Jacob T. and Bremer, Bennett J. and Romero, Philip A.},
  title = {Self-driving laboratories to autonomously navigate the protein fitness landscape},
  journal = {Nat. Chem. Eng.},
  volume = {1},
  pages = {97--107},
  year = {2024},
  doi = {10.1038/s44286-023-00002-4}
}

@article{wu2025,
  author = {Wu, Banghao and Zhong, Bozitao and Zheng, Lirong and Huang, Runye and Jiang, Shifeng and Li, Mingchen and Hong, Liang and Tan, Pan},
  title = {Harnessing protein language model for structure-based discovery of highly efficient and robust {PET} hydrolases},
  journal = {Nature Communications},
  volume = {16},
  pages = {6211},
  year = {2025},
  doi = {10.1038/s41467-025-61599-z}
}

@article{aledo2022,
  author = {Aledo, Juan Carlos},
  title = {renz: An R package for the analysis of enzyme kinetic data},
  journal = {BMC Bioinformatics},
  volume = {23},
  number = {1},
  pages = {182},
  year = {2022},
  doi = {10.1186/s12859-022-04729-4}
}

@article{ritz2005,
  author = {Ritz, Christian and Streibig, Jens C.},
  title = {Bioassay analysis using R},
  journal = {Journal of Statistical Software},
  volume = {12},
  number = {5},
  pages = {1--22},
  year = {2005},
  doi = {10.18637/jss.v012.i05}
}

@article{ritz2015,
  author = {Ritz, Christian and Baty, Florent and Streibig, Jens C. and Gerhard, Daniel},
  title = {Dose-Response Analysis Using {R}},
  journal = {PLOS ONE},
  volume = {10},
  number = {12},
  pages = {e0146021},
  year = {2015},
  doi = {10.1371/journal.pone.0146021}
}

@book{carroll1988,
  author = {Carroll, Raymond J. and Ruppert, David},
  title = {Transformation and Weighting in Regression},
  publisher = {Chapman and Hall},
  address = {New York},
  year = {1988}
}

@book{davidian1995,
  author = {Davidian, Marie and Giltinan, David M.},
  title = {Nonlinear Models for Repeated Measurement Data},
  publisher = {Chapman and Hall},
  address = {London},
  year = {1995}
}

@article{zeileis2006,
  author = {Zeileis, Achim},
  title = {Object-Oriented Computation of Sandwich Estimators},
  journal = {Journal of Statistical Software},
  volume = {16},
  number = {9},
  pages = {1--16},
  year = {2006},
  doi = {10.18637/jss.v016.i09}
}

@article{golub1973,
  author = {Golub, Gene H. and Pereyra, Victor},
  title = {The Differentiation of Pseudo-Inverses and Nonlinear Least Squares Problems Whose Variables Separate},
  journal = {SIAM Journal on Numerical Analysis},
  volume = {10},
  number = {2},
  pages = {413--432},
  year = {1973},
  doi = {10.1137/0710036}
}

@manual{nlme2026,
  author = {Pinheiro, Jose and Bates, Douglas and {R Core Team}},
  title = {nlme: Linear and Nonlinear Mixed Effects Models},
  year = {2026},
  note = {R package version 3.1-169 (used for the analyses)},
  doi = {10.32614/CRAN.package.nlme},
  url = {https://CRAN.R-project.org/package=nlme}
}
